# N. C. RANA: THE LIFE OF A `COMET' IN THE ASTROPHYSICAL WORLD

UTPAL MUKHOPADHYAY* AND SAIBAL RAY**

Abstract:
Narayan Chandra Rana, a person with extraordinary potential from a remote village of Bengal, India, came into the limelight of the international scientific world through his exceptional talent, zeal and courage. In his very short life-span, he excelled not only into various branches of astrophysics, but also took a leading role in science popularization, text book writing etc. In this paper, life and works of that budding scientist of India have been discussed from multifarious viewpoints.

**Key words:** Early life, Education, Academic position, Research works, N. C. Rana

## 1. Introduction

Narayan Chandra Rana, the eldest son of Rajendranath Rana and Nakfuri Rana, was born in the Sauri village of the undivided Midnapur (now in West Midnapur) district of Bengal, India on 10 October, 1954. Sister Bishnupriya and brother Sujan were younger than Narayan by two and ten years respectively. Rajendranath was an artisan of brass and bell metal and did brass-work in many temples of his own locality. He also maintained a *'pathsala'* (primary school run by a single person) in his own residence. Narayan Rana initially studied there. Afterwards, in March 1960 he took admission in class three in the primary section of Sauri Bholanath Vidyamandir. This institution was established on about 30 *bigha*$ land donated by Samarendra Chowdhury and was christened in memory of his father Bholanath Chowdhury, ex-*zamindar* (Land Lord) of Sauri. Headmaster of this primary school was amazed by Narayan's nice handwriting and his proficiency in mathematics. He was the person to remark for the first time that this boy would go a long way in his academic career.

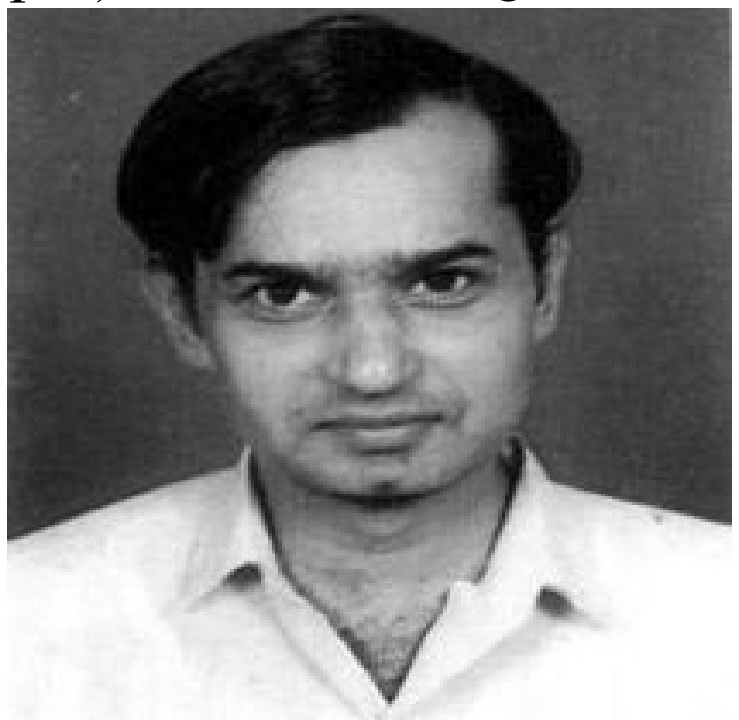

In the year 1963, Rana and his lifelong friend Gaur Kishore Raut (who later became a renowned doctor) entered in class five of the secondary section of Sauri Bholanath Vidyamandir. Rana and his classmates were third batch of students of this school. After getting promotion to class seven, Rana and Raut shifted to the school hostel on the same day. During Rana's school days, Srinibas Nanda, Patitpaban Bhowmik, Mohitosh Das, M.N. Lahiri, Gangesh Acharya, Chittaranjan Das etc. were his teachers. When Narayan was a student of class seven, his father died of pneumonia on 1 July, 1965 and financial condition of Rana's family became precarious.

*Satyabharati Vidyapith, Nabapally, Barasat, North 24 Parganas, Kolkata 700126, West Bengal, India; Email: utpalsbv@gmail.com
** Department of Physics, Government College of Engineering & Ceramic Technology, Kolkata 700010, West Bengal, India; E-mail: saibal@iucaa.ernet.in

$ *In West Bengal, the Bigha was standardized under British colonial rule at 1600 sq.yd (0.1338 hectare or 0.3306 acre); this is often interpreted as being 1/3 acre (it is precisely $^{40}/_{121}$ acre). In Metric units, a Bigha is hence 1333 m$^2$.*

But Rana overcame all obstacles by virtue of his exceptional talent and assistance of his teachers. Particularly, science teacher M.N. Lahiri (see *Appendix I*) was an infinite source of inspiration to Rana. After coming over to hostel, Rana had a close attachment with Mr. Lahiri. Lahiri's door was open to his students from the evening to 10 at night. Incidentally, Lahiri was an amateur sky watcher and Rana and Himanka Pal were his regular companions during night sky observation. Rana also received various science books from Mr. Lahiri. They were like close friends. Lahiri's ideals of life and love for sky became beacon of light for Rana in future. M.N. Lahiri commented about his beloved disciple that Rana never gave up before understanding any topic in his own way; he had no intention to escape from his studies. Lahiri also admitted that they (the teachers) had to keep themselves on their toes to answer the typical quarries of Rana. Mr. Lahiri himself had to study a lot and bring books for answering Rana's questions. While Rana was a student of class eight, by the assistance of teacher Mahitosh Das, he and Raut published a wall magazine entitled 'Kishore' (The Budding Young). In his school days, occasionally in absence of a number of teachers, Rana and Raut had to shoulder the responsibility of taking some of the classes. Moreover, Rana helped his classmates as well as students of lower classes when they faced any problem with their studies. As a student, Rana was exceptionally meritorious and always stood first from class five to class ten. Especially, his results in mathematics, science and geography were outstanding. According to his mathematics teacher Chittaranjan Das's version, after Test Examination of class ten, Rana requested him to prepare a difficult question paper for him, but the teacher realized that no question paper was difficult for Rana. In the year 1969, Rana stood second in the West Bengal School Final examination with letter marks in five subjects and was awarded National Scholarship. In that examination, Kartick Santra, Shyamal Sarkar and Sumitra Biswas secured first, third and fourth places respectively. His classmate and close friend Raut ranked 25$^{th}$ in the same examination. After hearing the news of Rana's success in the School Final Examination through radio broadcast in the evening on the date of the publication of the result, it was Raut who at first garlanded his bosom friend Rana. After passing School Final Examination, teachers M.N. Lahiri, Srinibas Nanda and the then Secretary of the school Nemaibabu shouldered almost all responsibility of Rana and Raut. They brought the two students to Mr. Lahiri's residence at New Alipore, Kolkata. From there, they were shifted to Ramkrishna Mission Vidyarthi Ashram at Belgharia. Meal, lodging, conveyance and even books and library facilities were arranged for Rana and Raut by their scholarship money. At that period, Vidyarthi Ashram was thronged by meritorious students of School Final and Higher Secondary classes. Monks of the mission, viz. Swami Dhyanatmananda, Swami Priti Maharaj, Swami Amalananda etc. were very affectionate to Rana. In the year 1970, Rana took admission in the Presidency College, Kolkata for studying in the Pre-University (P.U.) course. In the P.U. (Science) Final Examination, Rana and Raut were ranked 13$^{th}$ and 20$^{th}$ respectively.

## 2. Higher Education: Rana as a Student and as a Teacher

In 1971, Rana and Raut entered Presidency College as students of physics (honours). After six months, Raut was selected for M. B. B. S. course and he switched over to Nilratan Sarkar Medical College of Kolkata. But, Rana continued his studies at Presidency College and passed B. Sc. (honours) with flying colours in 1974. As a student of Presidency College, Narayan came in close contact with renowned physicist Prof. Amal Kumar Raychaudhuri (internationally famous for the equation which bears his name and has been widely used in cosmology describing the dynamics of the Universe) and became one of his dearest students. During 1974 to 1976, Rana studied his M. Sc. Course in physics. In M. Sc. Final examination, Rana stood second while Shyamal Sarkar, Achintya Pal and Chinmoy Ghosh secured first, third and fourth places respectively.

Since his college life, during rapid walking, Rana became out of breath very fast which was rather unusual for an young man. Initially he himself thought that it was caused by travelling in over-crowded train. But actually it was early symptoms of a heart disease. In medical terminology Rana's disease was 'Idiopathic Hypertrophic Sub-aeortic Stenosis with Complete Left Bundle Branch Block'. Rana's condition deteriorated gradually during his M. Sc. course. With the intervention of Monks of Belghoria Vidyarthi Mission, famous heart specialist Dr. R. N. Chatterjee started Rana's treatment. In the year 1977, Rana went to Mumbai for appearing in the Entrance Examination of Tata Institute of Fundamental Research (TIFR). Out of 256 candidates, 17 were selected and Rana stood first.

In TIFR, Rana started his research work under the supervision of famous cosmologist Prof. Jayant Vishnu Narlikar (born 19 July 1938). Title of Rana's research topic was 'Cosmic Microwave Radiation' and he received a scholarship of Rs. 650 per month. He completed his Ph.D. in 1981. The title of his thesis was 'An Investigation of the Properties of Inter galactic Dust'. It is interesting to note that he dedicated his thesis to Lord Ramakrishna, 'guru' of world famous monk Swami Vivekananda. His thesis was selected for 'Geeta Udgaonkar Award' from TIFR for the best thesis for the year 1983. In the same year, Rana received 'INSA Young Scientist Award' (usually awarded to scientists below 40 years for their outstanding research work) which was handed over by the then President of India Mr. Giani Jail Singh. But in the meantime, on 22nd October, 1978 pacemaker was planted in his heart and the entire expense was shouldered by the TIFR authority. Anyway, after completion of Doctoral work, Rana was appointed as a 'Permanent Research Fellow' with a monthly salary of Rs. 920. In the same year, Rana joined Durham University of England for Post Doctoral research work. At Durham he worked under the supervision of the famous astrophysicist Prof. Arnold Wolfendale (born 1927). The topic of his research was 'Chemical Evolution of the Galaxy'. Considering health condition of Rana, Prof. Wolfendale arranged for him a separate room and library in the ground floor. During his stay at Durham, Rana received two international awards, viz. 'Commonwealth Bursary Award' and 'SAARC Fellowship'. It is also to be mentioned that During Rana's stay at Durham, D.A. Wilkinson started his research for Ph.D. under Rana's supervision and completed his thesis by coming to Mumbai after his supervisor returned to India. Completing his Post Doctoral research work, Rana returned to India on 2nd October, 1985 and joined TIFR. His designation at TIFR was initially Fellow and then he was promoted to Reader post.

In 1986, Narlikar left TIFR and coming over to Pune he established his own research institute Inter University Centre for Astronomy and Astrophysics (IUCAA). On invitation from Narlikar, his scholar Rana joined IUCAA from TIFR as an Assistant Professor. In the year 1996, he was Associate Professor of Radio Astronomy and Astrophysics.

Table 1: *List of the scholars of N.C. Rana*

| Sl. No. | Name | Status | Duration |
|---|---|---|---|
| 1 | D.A. Wilkinson | Ph.D. | 1986-88 |
| 2 | Sarbani Basu | Ph.D. | 1990-93 |
| 3 | S.R. Pathak | Ph.D. | 1995 (Not completed) |
| 4 | R.K. Gulati | Post-Doc. | 1994 |

## 3. SCIENTIFIC WORKS: VERSATILE INTELECTUAL FACES OF RANA

N.C. Rana was altogether a theoretical astrophysicist. In his research career of seventeen years, spanning from late 70s to his premature death in 1996, his 58 research papers (single as well as joint) were published (see Reference of the present article and also the NASA- Arxiv and INSPIRE-Hep for a complete list). Those papers can be broadly classified into six major areas, viz. Astrochemistry, Stellar astrophysics, General Theory of Relativity and Cosmology, Celestial Mechanics, High Energy Astrophysics and Semi-popular Works (see Table 2 for getting a flavor of relevance/importance of Rana's scientific works).

### 3.1. Astrochemistry

During the 1960s, two cosmological theories were prevalent – Big Bang theory and Steady State theory. After the discovery of 3K cosmic microwave background radiation (CMBR), two schools of opinion emerged to explain that isotropic universal radiation. Proponents of Big Bang theory took CMBR as an observational evidence of the prediction of Gamow and his collaborators [1] while believers in Steady State theory engaged themselves for providing alternative explanations of CMBR. A series of three papers of Rana related to astrochemistry were in this direction.

In the first of these three [2], the role of natural graphite as a thermalizer for producing the observed 3K radiation was discussed. In fact, this work of Rana was an extension of the earlier works of Wickramsinghe et al. [3], Narlikar et al. [4], and Chitre and Narlikar [5]. It was shown by Rana that whiskers of long grains of natural graphite could not produce the observed 3K radiation. In the next paper [6] of this series, the role of absorption effect of X-ray and hard UV-ray were investigated for various cosmologies. Although a local density of the order of $10^{-34}$ $gmcm^{-3}$ was found to be inadequate for producing the background radiation, it was commented that the uncertainty exists regarding the absorption by X-rays up to Hubble radius. Although the previous papers could not provide an alternative explanation for production of CMBR, in the third paper in this line of work, Rana [7] showed that a density of $10^{-34}$ $gmcm^{-3}$ of pyrolytic graphite was sufficient for producing the observed background radiation.

Rana worked on the problem of estimation of molecular hydrogen in the Galaxy. In the first [8] of two papers, an empirical relationship of the form $\Psi_s \propto \sum^k_{H2}$ between the star formation rate and the surface density of molecular hydrogen has been derived. In the second paper [9], that relationship has been applied to estimate the radial distribution of molecular hydrogen in the Galaxy by using a compilation of star formation rates. The technique used by Rana and Wilkinson is different from the usual method of surveys. They compared their estimation with those obtained by CO surveys. This comparison clearly demonstrated the underestimation of the amount of molecular hydrogen found by CO surveys. By using Turner's model [10], Rana and Wilkinson [9] also derived the surface density of molecular hydrogen in the solar neighbourhood as $1.15 \pm 0.20\ M_\odot\ pc^{-2}$.

Amount of various chemical elements and their evolution with the passage of time was a favorite area of research for Prof. Rana. In a series of papers in the late 80s and early 90s, Rana made extensive investigation about chemical evolution of our galaxy, its relationship with star formation, dark matter and chemical evolution in the solar neighbourhood etc. In the first work of that series, in order to estimate star formation rate, Rana and Wilkinson [11] developed a model of chemical evolution of the Galaxy, in particular in the solar neighbourhood. It was shown that star formation rate was related to the surface density of molecular hydrogen. Moreover, the proposed model was used to address a number of important issues, viz. G-dwarf problem, the metallicity gradient, the stellar age-metallicity relation etc. It had been commented that distribution of molecular hydrogen as proposed by Bhat et al. [12,13] was preferable than that of Saunders et al. [14]. In the second paper of that series [15], some revisions of the first work [9] was done along with some other new findings. For instance, it was concluded that constant production of metallicity was preferred to the previous assumption of proportionality of yield with metallicity. It was found that 0.3 to 0.6 of mass fraction of dark matter was essential in the solar neighbourhood. It had been commented also that dark matter in the neighbourhood of the sun could be baryonic if Miller-Scalo initial mass function [16] was corrected at any end of the mass limits. Depending on the abundance of metals and molecular hydrogen, a new law of star formation rate $\Psi_s$ was proposed in the third paper [17] of that series. That new law was shown to be successful in explaining the constancy of $\Psi_s$ in the solar neighbourhood if the age-metallicity relation of Twarg [18] was taken into account. Moreover, it was found that with the assumption of that new law, a consistent model of the chemical evolution of galactic disc without infall [9] was possible. That law also explained the abundance of HI and lack of presence of much CO in sufficient number of stars in metal poor galaxies such as LMC and SMC [19].

How metals are distributed in dwarf stars lying in the solar neighbourhood had been investigated by Rana and Basu [20] in one of the papers of the same line of work. Using statistical method, it had also been found by the authors that the metallicity distribution obtained by them differed from that of Vilchez and Pagel [21] and Marsakov and Suchkov [22]. The calculated values of the parameters from the G-dwarf metallicity distribution were found to be consistent with the observed age-metallicity relationship. It is well known that mainly hydrogen and helium were produced primordially and other elements were synthesized within the stars. This means that evolution of chemical elements and compounds occur everywhere in the cosmos and hence the chemical composition of the space changes, both locally and globally, with subsequent evolution of the universe. In a major review work, Rana [23] took up the study of chemical evolution of our Milky Way galaxy. After presenting an overview of the present status of the topic, a model of phenomenological character was presented for addressing a number of important issues in that field. Moreover, relying on the current data set, a simplified model of chemical evolution was studied with reference to a single element iron.

As a continuation of their previous work [20], Basu and Rana [24] had investigated distribution of metals in G-dwarf stars in an inhomogeneous medium, the source of inhomogeneity being taken as birth of stars with a Gaussian distribution of metallicities in the logarithm of the metallicity around the mean metallicity of that particular epoch. It had been shown that tuning of parameters could produce plausible prediction for the age-metallicity relation. Moreover, the predicted shapes of the curves, in spite of being not completely satisfactory, were better in an inhomogeneous interstellar medium. Investigation of evolution of elements with time was a favorite topic of Prof. Rana. So, after working on chemical evolution of the Galaxy [15, 23], he went on to investigate the evolution of chemical elements in the solar neighbourhood [25]. Using a simple model, the chemical evolution and frequency distribution of a single element iron was observed. It had been shown that the calculated age-metallicity relation was consistent with the observed values. The Initial mass function (to be defined afterwards in the next subsection) of stars of masses exceeding 1.5 solar mass was found to obey a power law, the index being nearly - 1.67. Since Quasar absorption lines are good indicators of the chemical evolution of the universe up to $z \approx 4.0$, Khare and Rana [26] studied red-shift distribution of absorption lines of heavy elements in the QSO spectra for investigating the chemical evolution of high red-shift galaxies. It was found that amount of carbon increased by a factor of 5 to 20 during $z \approx 4$ to 2. The radius or the comoving number density of the galaxies increased up to $z \approx 2$ and then decreased. At $z = 4$, almost entire mass of the galaxy was concentrated in the halo in the form of gas. The observed hydrogen column density for the Lyman limit systems was found to be consistent with the theoretical prediction from the model. As a more general work of previous two works [15, 23], Basu and Rana [27] investigated chemical evolution of the Galaxy when the star formation rate was nearly constant. By assuming a star formation rate ($\Psi$) of the form $\Psi \propto \sum^{\alpha}_{g} z^{\beta}$, the authors first investigated the solar neighbourhood and then extended their results to the entire Galaxy. It was shown that a better result than Rana and Basu [24] could be obtained if $\alpha = 1.15 \pm 0.05$ and $\beta = 1.25 \pm 0.05$. Moreover, a better result could be obtained if one assumes a variation of star formation rate across the face of the Galaxy by about 10%. During star formation, first deuterium gets destructed and hence relative amount of deuterium decreases during evolution of the Universe. This destruction of deuterium is known as astration of deuterium. Destruction of deuterium within the stars of the Galactic disc near the sun was investigated by Rana and Kousalya [28]. It was shown that for a closed model of chemical evolution, the astration of deuterium over a period of 13 Gyr corresponds to a factor of 2. In another paper, Rana and Jana [29] worked on the distribution of metals in F-dwarf stars residing in the solar neighbourhood. By choosing a volume limited sample of stars with low radial velocity ($|v_z| \leq 40$ kms$^{-1}$ at $z \leq 100$ pc) from Knude [30] and taking into consideration the possible chemical inhomogeneity, it was shown that the expected value of Fe/H was consistent with the calculated ones [31].

As a continuation of their previous work [26], Khare and Rana [32] investigated the chemical evolution of galaxies at $z = 2$ to 4 using the spectra of quasars. By assuming the galactic haloes as spherical with radius about 50 kpc and assuming also a diffuse UV radiation background which increases as $(1 + z)^2$ up to $z = 2.0$ and as $(1 + z)^{0.2}$ at $z > 2$ [33], it was shown that the amount of gas (in terms of mass) was at least three times in the past at $z = 4$ and the amount of carbon has increased 5 – 20 times from $z = 4$ to 2. Pathak and Rana [34] used observational luminosity function and star formation rate for white dwarfs for finding the time required for complete crystallization of the oldest white dwarf stars. It was shown that more than 10 billion years are required for complete crystallization of a typical white dwarf star.

### 3.2. Stellar Astrophysics

An important parameter in the field of stellar astrophysics is the 'Initial Mass Function' (IMF) of stars which is an empirical function that describes the mass distribution of a population of stars in terms of its theoretical initial mass. Since path of stellar evolution depends on stellar mass, so it is an important concept to the astronomers. Salpeter [35] first obtained a formula describing the IMF of stars heavier than sun. Afterwards Miller and Scalo [36], Kroupa [37] and Chabrier [38] had derived IMF for stars of different mass range.

Using a law of star formation derived by Rana and Wilkinson [9] and some recent data on luminosity function and scale heights of main sequence stars, Rana [39] made an investigation about the mass function of those stars. A comparison of the obtained results with those of Scalo [40] and Larson [41] showed that classical views, viz. near constancy of star formation rate, an IMF as a simple power law, no exotic amount of dark remnants or brown dwarfs etc. were still useful enough for offering a consistent view of the chemical evolution of the region near the sun.

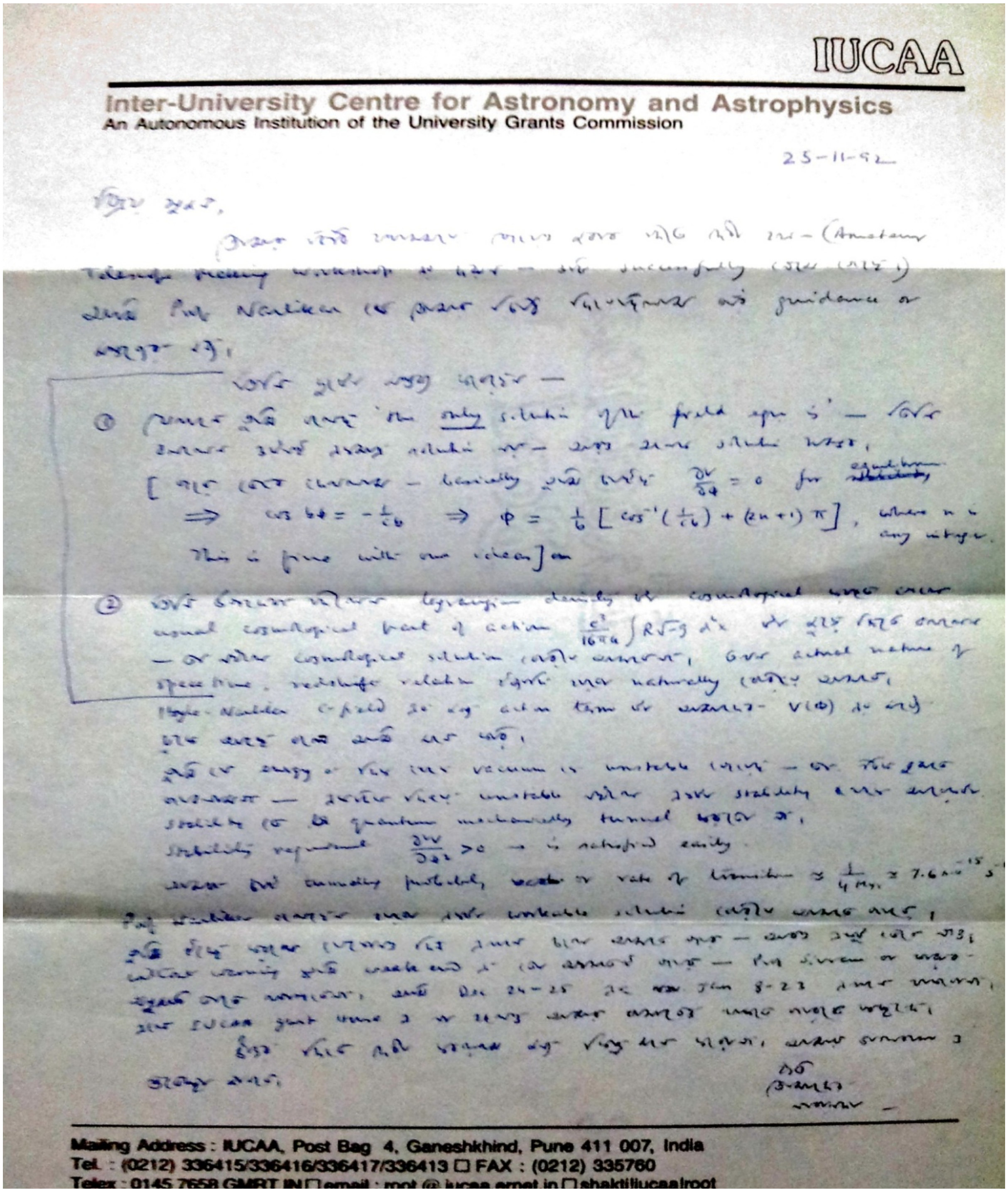
IUCAA

Inter-University Centre for Astronomy and Astrophysics

An Autonomous Institution of the University Grants Commission

25-11-92

Mailing Address : IUCAA, Post Bag 4, Ganeshkhind, Pune 411 007, India

Tel. : (0212) 336415/336416/336417/336413 □ FAX : (0212) 335760

Photo 1*: A hand written letter in Bengali by Prof. N.C. Rana to Prof. Sujan Sengupta, I.I.A., Bangalore on 25.11.92 regarding conversation with and suggestion by Prof. Narlikar on some scientific issues of Prof. Sengupta* [In courtesy of Prof. Sujan Sengupta, IISc, Bangalore].

It is well known that star formation rate in spiral galaxies depends on the presence of molecular hydrogen. Rana and Wilkinson [42] proposed a law of star formation in our Galaxy. In that work, Rana and Wilkinson [42] showed that the ratio of molecular hydrogen and the total amount of gas might be related to the metallicity for a number of nearby spiral galaxies. It was also possible to derive an empirical star formation law that might be helpful for investigating the history of chemical evolution of the spiral galaxies mentioned earlier. Moreover, it was observed that the law of star formation derived by Rana and Wilkinson [42] for spiral galaxies could be used for nearby spirals as well.

In another work by Rana [43], observing the luminosity distribution and the theoretical cooling curves of white dwarfs of mass about 0.6 solar mass which were formed as final stages of intermediate and low mass stars, the rate of formation of those white dwarfs as well as their progenitors in the solar neighbourhood had been estimated as a function of time. That estimation function showed a nearly constant star formation rate for last 10 – 12 Gyr and the observed number density of the local white dwarfs was consistent with the expected value derived from the mass function of stars in the solar neighbourhood. In another work, Rana [44] , using data on the luminosity function and lifetime of the main sequence stars, derived an IMF of main sequence stars in the solar neighbourhood. The proposed IMF showed multimodal star formation and was helpful in settling the missing mass issue satisfactorily. Using the data of Pagel [45] and Solderblom et al. [46], Rana and Basu [47] derived age-metallicity relation (AMR) for sun like dwarf stars in the solar neighbourhood. Choosing a simple closed model, the evolution of a single element iron was shown to be in agreement with the AMR obtained by this work. Moreover an IMF of local stars was derived by using the star formation rate in the proposed model as well as that obtained by Soderblom et al. [46]. This IMF was successful in nullifying the presence of hidden mass in the local disk. In a subsequent work, Basu and Rana [48] derived the present day mass function (PDMF) in the solar vicinity after proper correction for the effects of unresolved multiple stellar systems. This corrected PDMF gave the surface mass density as 39.5 $M_\odot pc^{-2}$ as compared o 31.3 $M_\odot pc^{-2}$ for uncorrected PDMF. This new PDMF was used for deriving the IMF. It was found that for stellar masses between 1.4 $M_\odot$ to 6.5 $M_\odot$, the index of the power law for the IMF was – 1.56 ± 0.05 whereas better result could be obtained for the index – 1.67 for more massive stars. It was also shown that the reciprocal of the time constant should lie in the range – 1.0 to + 1.0 as restricted by the formation rate of neutron stars and white dwarfs.

As a continuation of this work, Basu and Rana [49] corrected the IMF by taking into account the multiplicity of stars in stellar systems. In fact, the IMF was derived from the PDMF based on observed luminosity function of stars. Three cases were considered, viz. (1) the primary and the secondary with equal masses, (ii) the mass of the primary was twice that of secondary and (iii) the secondary to primary mass ratio increased from 0.25 for primary stars of mass less than 1 $M_\odot$ to 1.0 for primary masses greater than 10 $M_\odot$. The corrected PDMF showed that the surface mass density of stars in the solar neighbourhood was 41.3 $M_\odot pc^{-2}$ as compared to 32.7 $M_\odot pc^{-2}$ for uncorrected PDMF. This increase in the estimation of mass was supposed to do away with the concept of dark matter in the solar neighbourhood. A number of arches have been observed in the Eta Carinae nebula. Without taking it granted as chance alignment, Rana and Gajria [50] investigated the physical cause of the existence of a large number of arches in the Eta Carinae. After studying 96 such arches, the authors came to the conclusion that the arches were results of supernova induced star formation. Incidentally Rana and Gajria [50] found that the single, binary, triple and quadruple stellar systems in the Eta Carinae were in the ratio 52: 35: 11:

2 which is similar to 51: 39: 8: 2 as found by Duquennoy [51]. Srivastva, Gulati and Rana [52] used the data of Andersen [53] on the mass, absolute visual magnitude, effective surface temperature, bolometric corrections etc. for making a new calibration for those parameters. The latest evolutionary models was compared with the empirical data for justifying the consistency of SSMM models [54]. Low mass stars residing in the galactic halo (called MACHO) are considered as a possible candidate for dark matter. Shortly after the discovery of MACHOs, Rana and Chandola [55] calculated the possible increase in mass-luminosity ratio (M/L) by taking into account the MACHOs and using the input data from the latest PDMF of stars. It was shown that if any system contained a large number of stars of mass about 0.1 $M_\odot$ or so, then the M/L ratio could go up to an order of $10^3$. In another work, Rana and Chandola [56] worked on mass functions of stars in the solar neighbourhood. Rana [57] worked on the scale heights of low mass stars. Using the observed luminosity function of the local white dwarfs and the theoretical cooling rates of a typical white dwarf, it was shown that the formation rate of white dwarfs was nearly constant and that was about three times lower than their progenitors.

### 3.3. General Theory of Relativity and Cosmology

In his first work related to general theory of relativity (GTR) [58], Rana (along with Narlikar) showed that 3K background radiation could be explained in terms of primordial black holes. In the next paper of that series, Datta and Rana [59] showed that finite range gravitation was inadequate to provide a consistent cosmological solution of the field equation and its prediction regarding the red-shift formula and the value of the Hubble constant were inconsistent with the observational result. Moreover, the domain of the scale factor R was near unity and hence the idea of a Big Bang could not be accommodated in the theoretical framework. So it was necessary to invoke other astrophysical process for explaining the CMBR. In another work related to the explanation of the microwave background radiation, Narlikar and Rana [60] showed that cosmological theories with varying gravitational constant could support the observational data better than that in the Friedmann models. In order to overcome the difficulties associated with the standard interpretation of the CMBR, Rana [61] showed that a large number of stellar type nucleosynthesis could provide CMBR like radiation and radiation by ambient intergalactic dust medium could give sufficient thermalisation so as to produce the same spectral signature as CMBR. The model presented there was shown to be consistent in terms of helium abundance, the baryonic matter density of the universe, the degree of isotropy of CMBR etc. Stecker [62,63] demonstrated some inconsistencies between the theoretical calculations and observational findings regarding the primordial abundances of helium and deuterium. Rana [64] extended Stecker's work farther to show that in the standard BBN calculations, no single value of $\eta$ (the ratio of baryon density to photon density) could produce the exact abundances of helium and deuterium. However, slight neutrino degeneracy could produce a plausible solution for $\eta$. As a subsequent work of the previous one in the G-varying cosmology [60], Narlikar and Rana [65] demonstrated that the theoretical curve of the CMBR could show a better fit in a G-varying cosmology both at long and short wavelengths. In another work, Rana [66] described the early universe with special emphasis to the synthesis of helium and deuterium. Cosmic rays whose origin is still unknown, has some important cosmological significance because cosmic rays played important role in galaxy formation, low energy antiprotons indicated the existence of both galaxies and antigalaxies etc. Rana and Wolfendale [67] investigated the relationship between cosmic rays and cosmology. Some of the cosmological implications, viz. the origin of the extragalactic gamma ray background, the problem of generating particles with highest energies, the importance of the excess of low energy antiprotons in the cosmic ray beam, the role of cosmic rays in galactic dynamics, particularly galactic collapse etc. were also discussed. The importance of the presence of cosmic rays in the early universe was also demonstrated. Roy and Rana [68] used the harmonic gauge, Chandrasekhar gauge and a composite Chandrasekhar gauge for finding the metric as well as equations of motion in the first order post-Newtonian approximation for EIH (Einstein, Infeld and Hoffman) equations of motion for N finite mass points interacting with each other through gravitational force. It was observed that although the equations of motions were all identical, the solutions obtained were all different. It was also found that the accuracy of the second order contribution for acceleration of any planet would be better than one part in $10^{17}$ if one used the Chandrasekhar gauge and his formalism were used. In another paper, Rana [69] worked on the early universe and the Big Bang nucleosynthesis from the cosmological point of view. In a subsequent work, Roy and Rana [70] critically examined the self-consistency of EIH equations of motion in a special case for three bodies with two bodies very close to each other while the third one was far away from them. Rana and Mitra [71] studied the effect of neutrino heating in the early universe. Following the line of work of Herrera and Hacyan [72,73] and assuming neutrinos as massless particles, it was shown that electron neutrinos decoupled at $1.5 \times 10^{10}$ K and other families of neutrinos decoupled at $2.5 \times 10^{10}$ K. The effect of that heating should reduce the amount of primordial helium in the standard hot Big Bang model by nearly 0.003. As a continuation of their previous work [70], Roy and Rana [74] critically analyzed the situation when a static spherically symmetric body would not appear as a point mass in the light of the Chandrasekhar gauge. The result obtained was compared with the Schwarzschild solution and certain problems were pointed out.

Table 2: *List of the research papers by N.C. Rana to get a flavor of relevance/importance of his works through citations*

| Sl. No. | Authors | Journal reference | No. of citation |
|---|---|---|---|
| 1 | N.C. Rana | 1979Ap&SS..66..173R | 13 |
| 2 | N.C. Rana | 1980Ap&SS..67..201R | 2 |
| 3 | J.V. Narlikar and N.C. Rana | 1980PhLA...77..219N | 10 |
| 4 | N.C. Rana | 1980Ap&SS..71..123R | 10 |
| 5 | N.C. Rana | 1981MNRAS.197.1125R | 10 |
| 6 | N.C. Rana | 1982PhRvL..48..209R | 27 |
| 7 | J.V. Narlikar and N.C. Rana | 1983PhLA...99...75N | 1 |
| 8 | N.C. Rana | 1984NCimB..84...53R | 4 |
| 9 | N.C. Rana and A.W. Wolfendale | 1984VA.....27..199R | 1 |
| 10 | N.C. Rana et al. | 1984A&A...141..394R | 12 |
| 11 | J.V. Narlikar and N.C. Rana | 1985MNRAS.213..657N | 5 |
| 12 | N.C. Rana and D.A. Wilkinson | 1986MNRAS.218..497R | 57 |
| 13 | N.C. Rana and D.A. Wilkinson | 1986MNRAS.218..721R | 8 |
| 14 | N.C. Rana and D.A. Wilkinson | 1986NCimC...9..714R | 1 |
| 15 | N.C. Rana and D.A. Wilkinson | 1987IAUS..120..323R | 1 |
| 16 | N.C. Rana and D.A. Wilkinson | 1987MNRAS.226..395R | 9 |
| 17 | N.C. Rana | 1987A&A...181..195R | 2 |
| 18 | N.C. Rana | 1987A&A...184..104R | 72 |
| 19 | N.C. Rana and D.A. Wilkinson | 1988MNRAS.231..509R | 11 |
| 20 | N.C. Rana | 1989LNP...328..152R | 2 |
| 21 | N.C. Rana | 1990ASSL..162..381R | 2 |
| 22 | N.C. Rana | 1990Ap&SS.163..229R | 5 |
| 23 | R. Roy and N.C. Rana | 1990Ap&SS.167..125R | 1 |
| 24 | N.C. Rana and S. Basu | 1990Ap&SS.168..317R | 14 |
| 25 | P. Bhattacharjee and N.C. Rana | 1990PhLB..246..365B | 62 |
| 26 | R. Roy and N.C. Rana | 1990JApA...11..291R | 1 |
| 27 | N.C. Rana and B. Mitra | 1991PhRvD..44..393R | 31 |
| 28 | S. Basu and N.C. Rana | 1992ApJ...393..373B | 51 |
| 29 | S. Basu and N.C. Rana | 1992Ap&SS.196....1B | 9 |
| 30 | N.C. Rana and S. Basu | 1992A&A...265..499R | 54 |
| 31 | P. Khare and N.C. Rana | 1993JApA...14...83K | 5 |
| 32 | R. Roy and N.C. Rana | 1993Ap&SS.207..145R | 1 |
| 33 | S. Basu and N.C. Rana | 1993ApJ...417..145B | 3 |
| 34 | N. Srivastva, R.K. Gulati and N.C. Rana | 1994JApA...15..187S | 1 |
| 35 | S. Chakrabarty and N.C. Rana | 1994JPhG...20L.117C | 1 |
| 36 | N.C. Rana and M. Chandola | 1997seas.conf..307R | 1 |
| Total citations | | | 497 |

### 3.4. High Energy Astrophysics

High energy phenomena going on in the cosmos, viz. gamma ray bursts, cosmic ray shower, hard x-ray emission etc. are topics of astrophysical interest to the scientists. In his research career, Professor Rana touched upon this branch of astrophysics and published a few papers on it.

In his first work in this line, keeping in view the then discovery of sources of high energy cosmic gamma rays like Cygnus X-3, Rana et al. [75] determined the order of magnitude of hard X-ray emission due to synchrotron radiation at high Galactic latitudes, the physical process involved being the collision of cosmic high energy gamma rays (energy > $10^{15}$) with the CMBR. It was shown that only one source of gamma ray like Cygnus X-3 was sufficient to produce X-ray with energy above nearly 100 Kev. As a sequel of this work, Rana et al. [76] examined a number of ultra high energy (UHE) gamma ray sources and inferred that flux of hard x-rays should be latitude independent. Origin as well as propagation of cosmic ray particles was also discussed from theoretical point of view. In another work, Bhattacharjee and Rana [77] investigated the influence of collapse or multiple self–interactions of a particular class of closed cosmic string loops for the generation of UHE (energy> $10^{18}$) protons as expected in the present epoch. That expected flux was compared with the observed flux of UHE cosmic rays. In a work related to solar physics, Chakrabarty and Rana [78], assuming a partial distributions of electrons of non-thermal origin, considered the electron capture rates of the reactions $^{3}He\ (e^{-}, \nu_e)^{3}H$, $^{1}H\ (p\bar{e}, \nu_e)^{2}H$ and $^{7}Be\ (\bar{e}, \nu_e\gamma)^{7}Li$. It was shown that the reactions having negative Q-values were drastically modified while the reactions with positive Q-values were modified by 30-40%. Particularly in the case of the reaction $^{3}He\ (e^{-}, \nu_e)^{3}H$, the process was found to be a cycle $^{3}H \leftrightarrow {}^{3}He$ for temperature less than $10^{8}$ K. The significance of $^{8}Be$ solar neutrinos was also discussed by the authors.

### 3.5. Celestial Mechanics

Rana made some contributions in the field of celestial mechanics also. Applying the technique of numerical integration of the Newton's equations of motions and gravitation, Narlikar and Rana [79] calculated the advancement of Mercury's perihelion. It was shown that the rate of advancement fluctuates around a central steady value of 528.95 $arcseccy^{-1}$ extending over a period of five centuries. A comparison with the observed value showed a discrepancy of nearly 2.3 $arcseccy^{-1}$ with the general relativistic correction. That result prompted the authors to suggest another reason like solar oblateness for explaining the error. As a continuation of this work, Rana [80] numerically calculated the motions of the node and the perihelion of Mercury and compared the result obtained with those found analytically and observationally. The differences with the result of analytical method were found to be negligible. It was found that the geocentric angular data of low precision (± 1) could not explain the apse rate better than 3 arcsec per century due to the geometry of the orbits. It was commented that an accuracy of apse rate of 0.1 arcsec per century could be obtained by only very accurate range measurements (± 1 km or less) on carefully selected portions of the orbit over a long enough period of about one or two decades.

In another work, Mitra and Rana [81] analyzed both yearly (1620-1987) and daily (1977-1985) time series data for searching (if there is any) periodicities of the variation of the length of the day. Various techniques like FFT technique, Scargle algorithm, Burg's method etc. were used to show that annual, bi-annual and monthly variations exist beyond doubt while long term periodicities of 184, 61, 40.9 years were also possible. But those periodicities were probably unrelated to the equilibrium tidal effects of the Sun and the Moon. The amplitude of the 18.6 year periodicity was also calculated from the daily time series of the length of the day. In a subsequent paper, Rana and Mitra [82] analyzed the astronomical data of Stephenson and Morrison [83] for the period 1860-1980 on the rate of lengthening of day and the tidal and non-tidal parts were recalculated independently. The overall agreement was found to be marginal compared to the amount of disagreement shown by Peltier [84] and Stephenson and Said [85]. In another work related to the interconnection of solar oscillation and mass extinction, Das and Rana [86] calculated the period of Solar oscillation about the Galactic plane using the then latest model of the vertical structure of the Galactic disc in the neighbourhood of the Sun and showed that a quarter of the period of oscillation matched with periodicity of mass extinction.

## 4. Rana as a Science Writer and Popular Science Worker

Prof. Rana published over 50 research papers and hundred articles in various journals and magazines [87]. A glance at the topics of those papers reveal that he worked in several areas, viz. astrochemistry, cosmology, stellar astrophysics, celestial mechanics, high energy astrophysics, history of astronomy etc. He had also written a number of popular science articles, both in English and Bengali language. Besides these research and popular articles, Prof. Rana had authored five books – (1) Classical Mechanics (with P. S. Joga) [88], (2) Our Solar System (with A. W. Joshi) [89], (3) Night Fall on a Sunny Morning (with N. Vayada) [90], (4) Myths and Legends Related to Eclipses [91] and (5) Observer's Planner '97. Among those five books (see the list in the Table 3), the second one is widely used as a text book at college and university level and is really 'classical' in the true sense of the word. Prof. Rana dedicated this book to his respected teacher Prof. A. K. Raychaudhuri.

Table 3: *A list of Books written by N.C. Rana*

| Sl. No. | Authors | Title of the book | Publisher |
|---|---|---|---|
| 1 | N.C. Rana and P. S. Joga | Classical Mechanics | Tata McGraw-Hill (1991) |
| 2 | A W. Joshi and N C. Rana | Our Solar System | Wiley Eastern Limited (1992) |
| 3 | N. Vayada and N C. Rana | Night Fall on a Sunny Morning | Confederation of Indian Amateur Astronomers (1994) |
| 4 | N.C. Rana | Myths and Legends Related to Eclipses | Vigyan Prasar, New Delhi (1995) |
| | N.C. Rana | Observer's Planner '97 | |

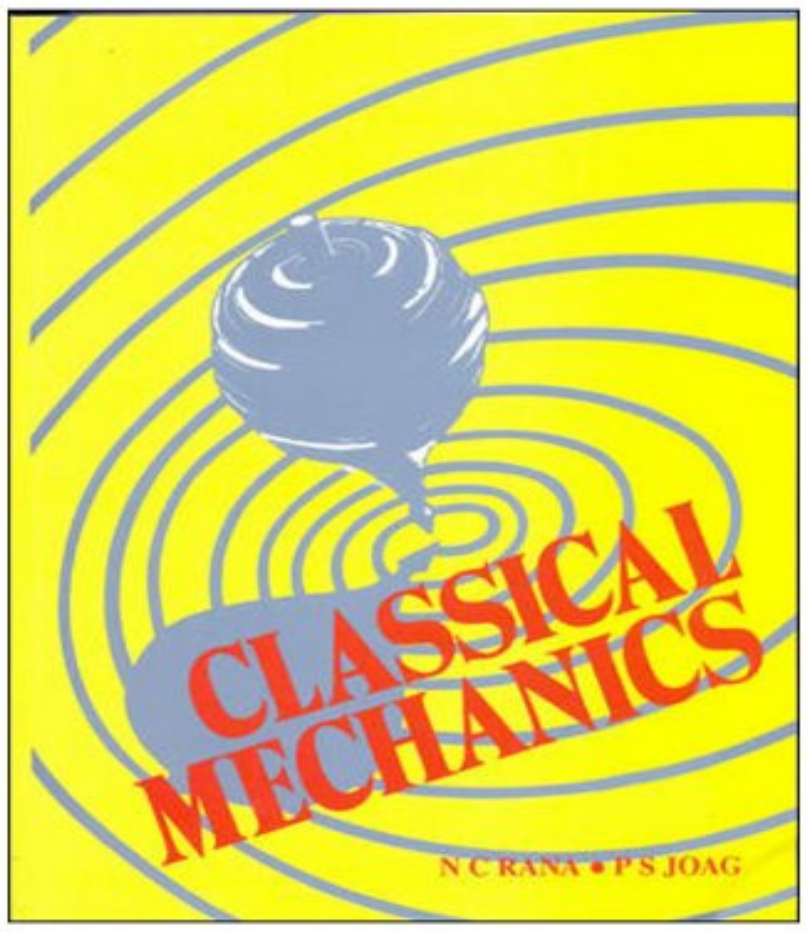

Photo 2: *The coverpage of Classical Mechanics*

In 1995 Rana received 'Best Teacher Award' from Pune. He was the main architect of popularization of amateur astronomy and was Indian representative of International Astronomical Union (Teaching of Astronomy). He was instrumental in the formation of 'Confederation of Indian Amateur Astronomers' (CIAA), a platform for interaction of amateur astronomers of India and was its Chairman. As far as IUCAA's first Outreach Programme is concerned Rana (as a chairperson), J.V. Narlikar, Ranjan Gupta and Arvind Paranjape (as a convener) were the first Outreach Committee members of it. However, later on other members were changed by rotation but Rana and Paranjape both remained in their respective positions [92]. It was an anecdote by Prof. Narlikar, in his 25th IUCAA Memorial Lecture delivered on 29 December 2013, that the dome in front of the IUCAA canteen was the brain-child of Rana. It was built on 29 December 1988, just after a few years of IUCAA's formation [93].

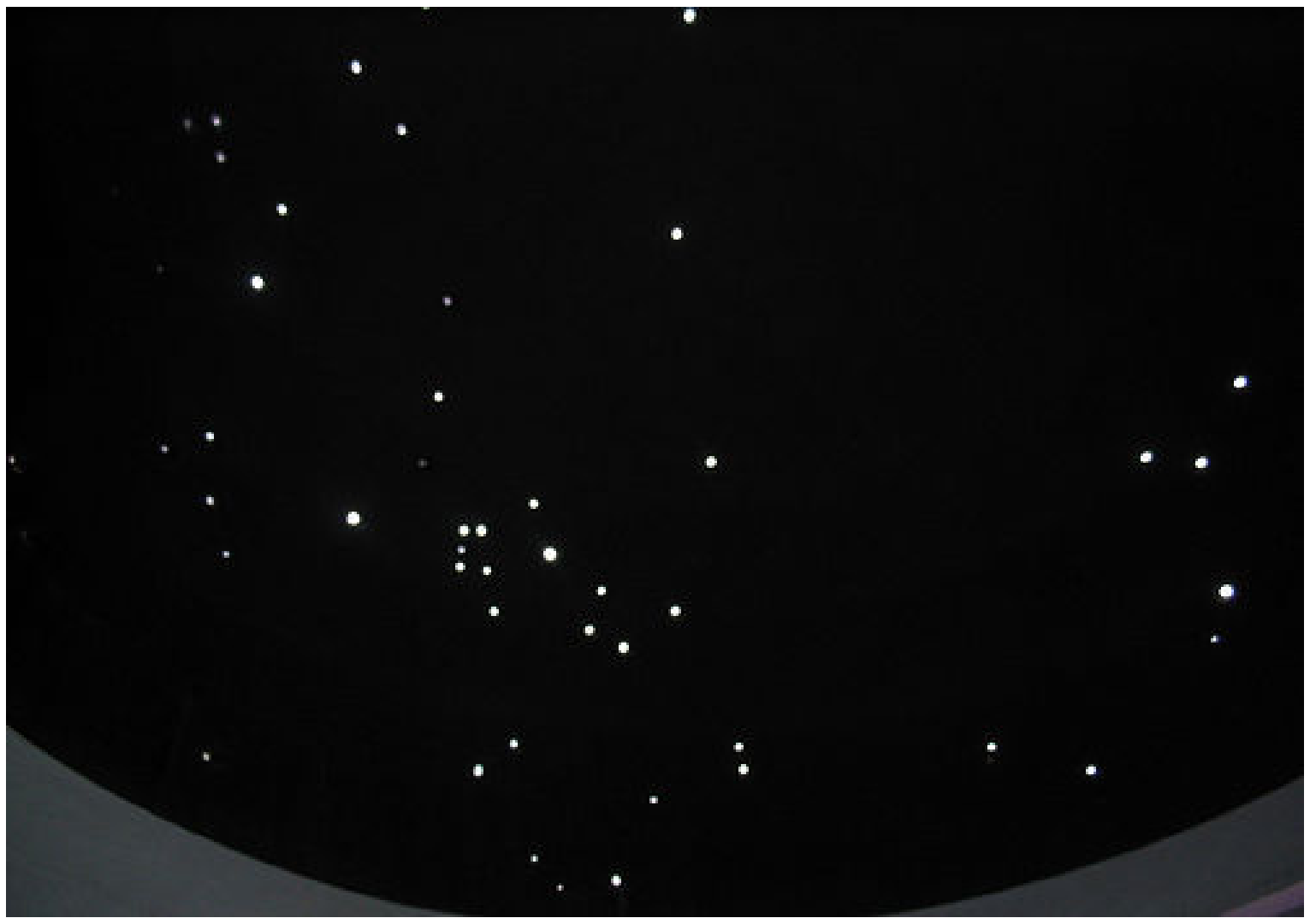

Photo 3: *The photograph of the perforated dome (inside view) through which night stars are visible* [In courtesy of Samir Dhurde, IUCAA, Pune]

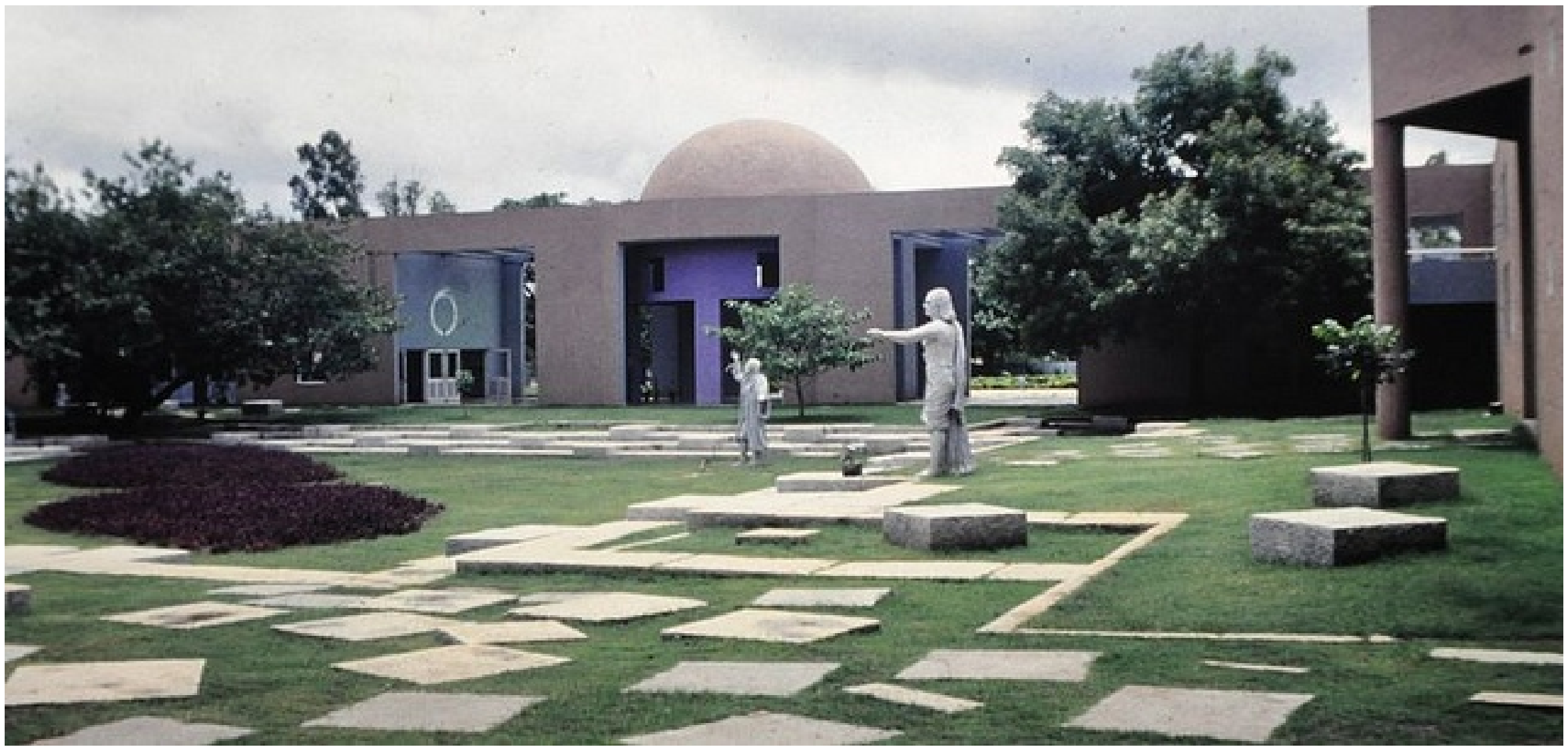

Photo 4: *The photograph of the dome (global view*) [In courtesy of Samir Dhurde, IUCAA, Pune]

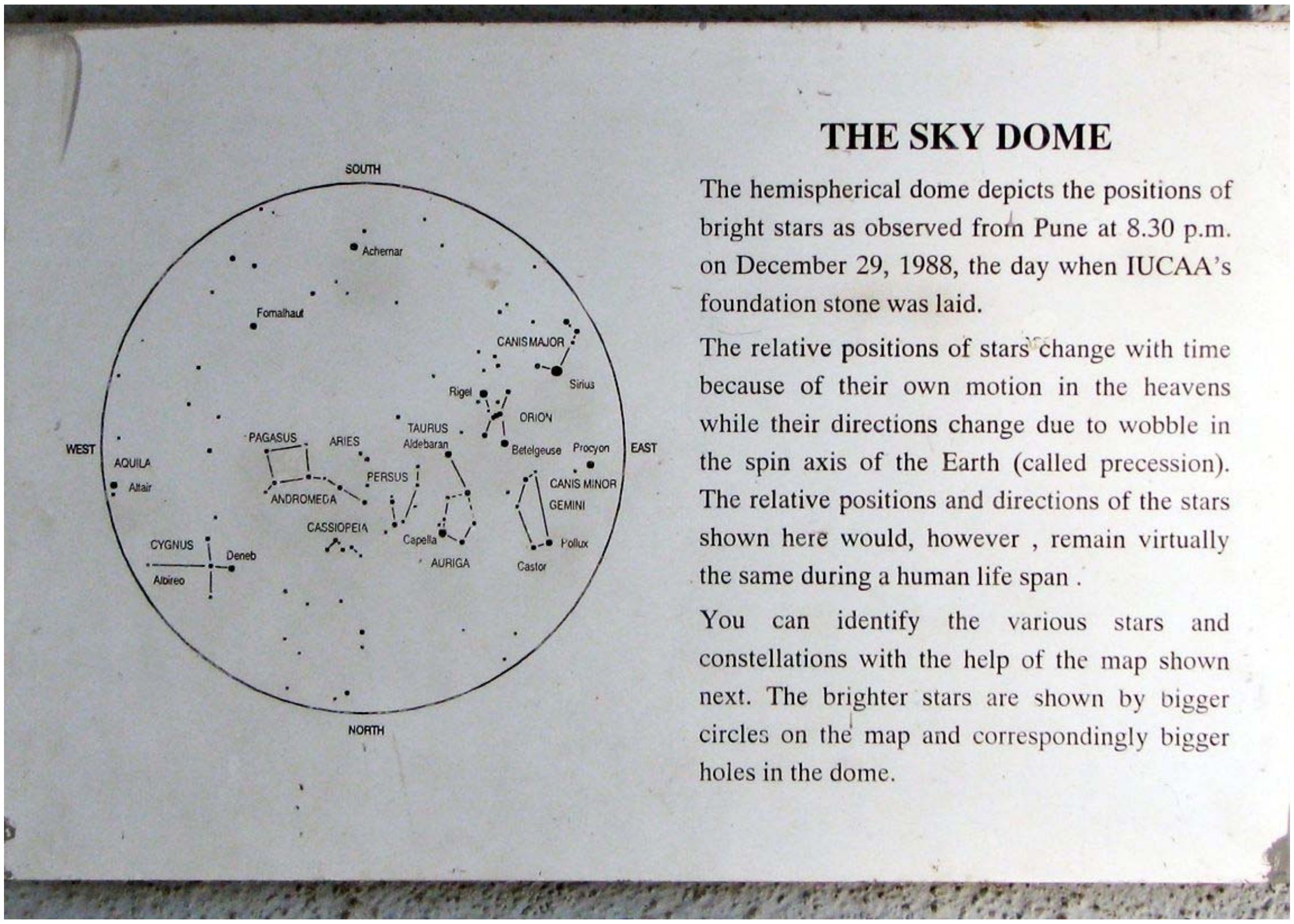

Photo 5: *The photograph of the plaque on the inside wall of the dome* [In courtesy of Samir Dhurde, IUCAA, Pune]

Some rare photographs of Prof. N.C. Rana in different role: as popular science activist, as astronomy teacher and as Faculty Member (Photo 6 - 16) *[In courtesy of IUCAA Library, Pune]*

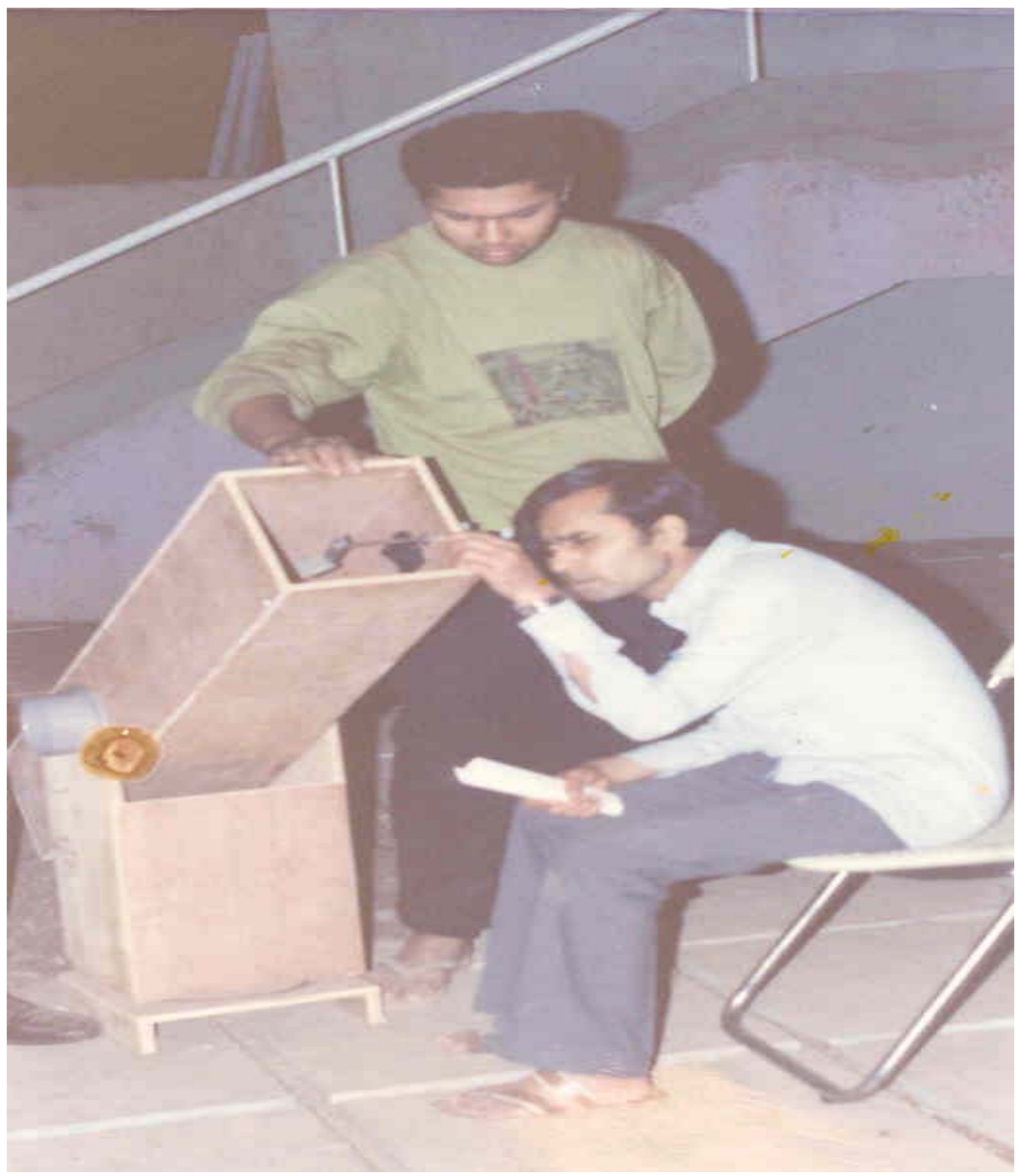

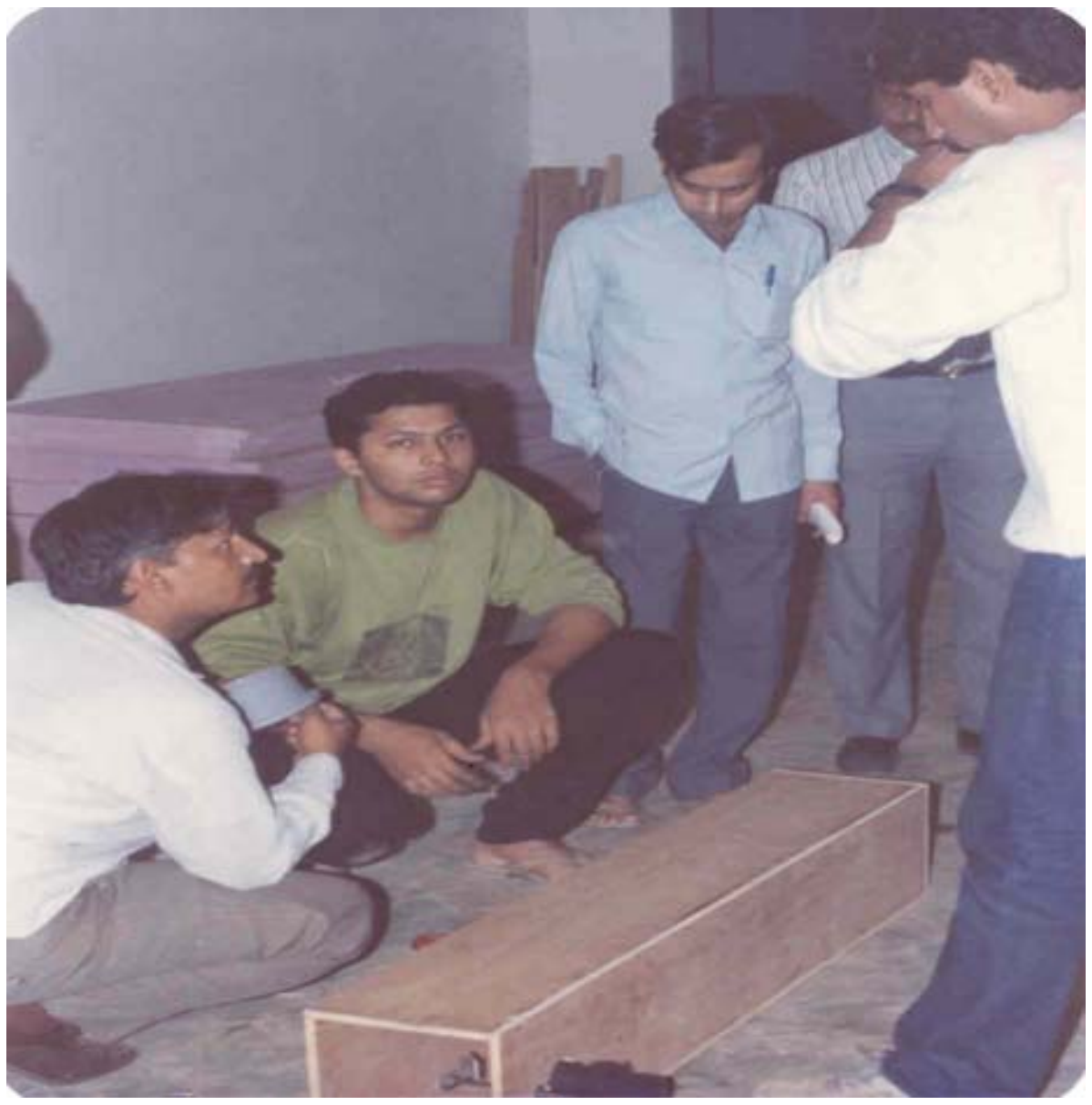

Photo 6: Prof. N.C. Rana critically examines the collimation of one of the 6 inch telescopes in a Telescope Making Workshop (left panel). Photo 7: A 6 inch telescope made at the Telescope Making Workshop (right panel).

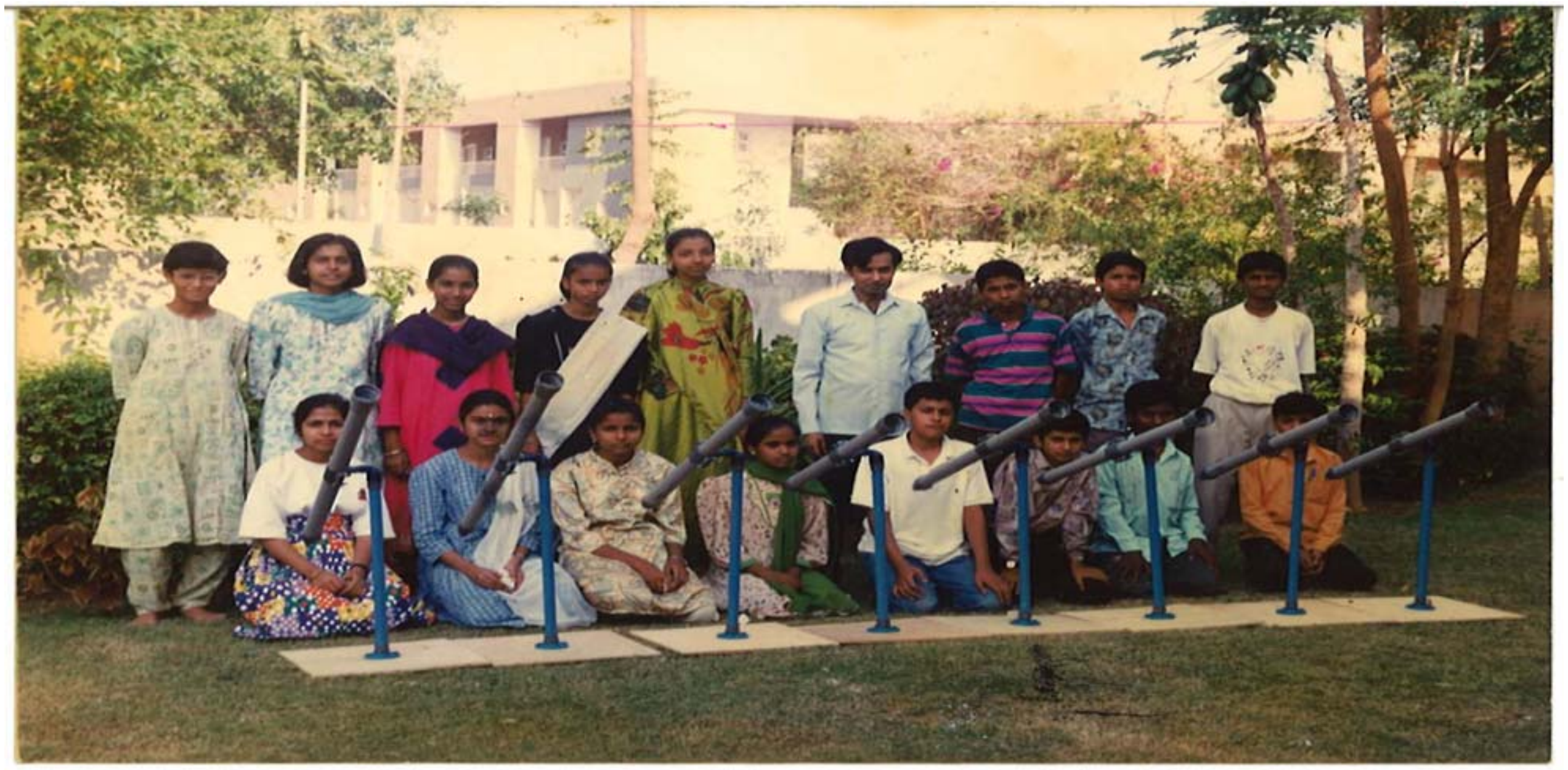

Photo 8: Summer Programme for School Children with Professor N.C. Rana, at IUCAA, Pune

Photo 9: Prof. Rana with other resource persons of the Regional School on Introductory Astronomy at Jalana, Maharashtra, Sept. 7-12, 1992. In his right Mr. Arvind Paranjape (IUCAA) whereas on the left Prof. Ranjan Gupta (IUCAA) and Prof. A.K. Sen (Assam University).

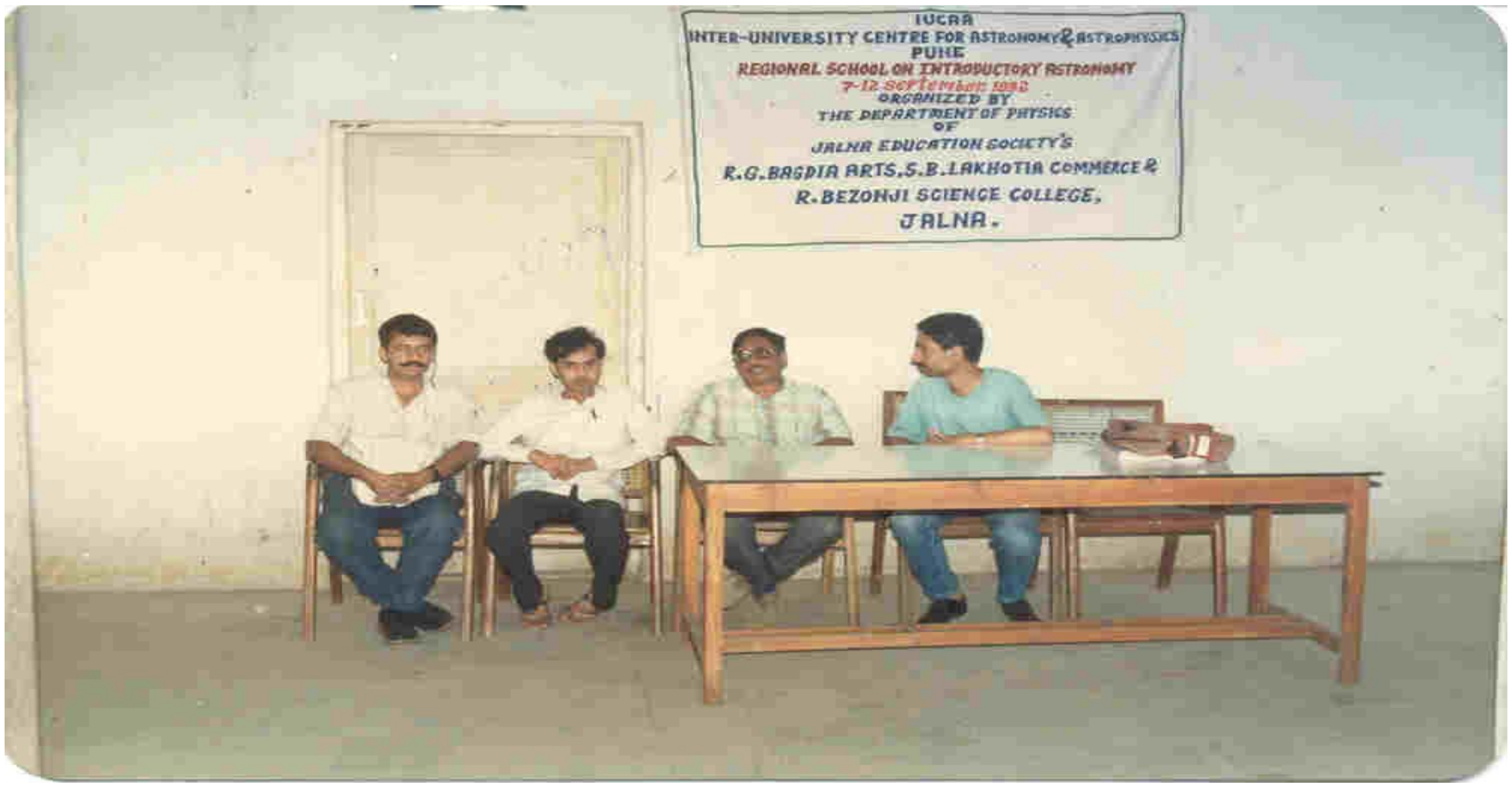

Photo 10: Resource persons along with the participants of the Regional School on Introductory Astronomy at Jalana, Maharashtra, Sept. 7-12, 1992 where Rana is seen standing in the 1st row (3rd from the left)

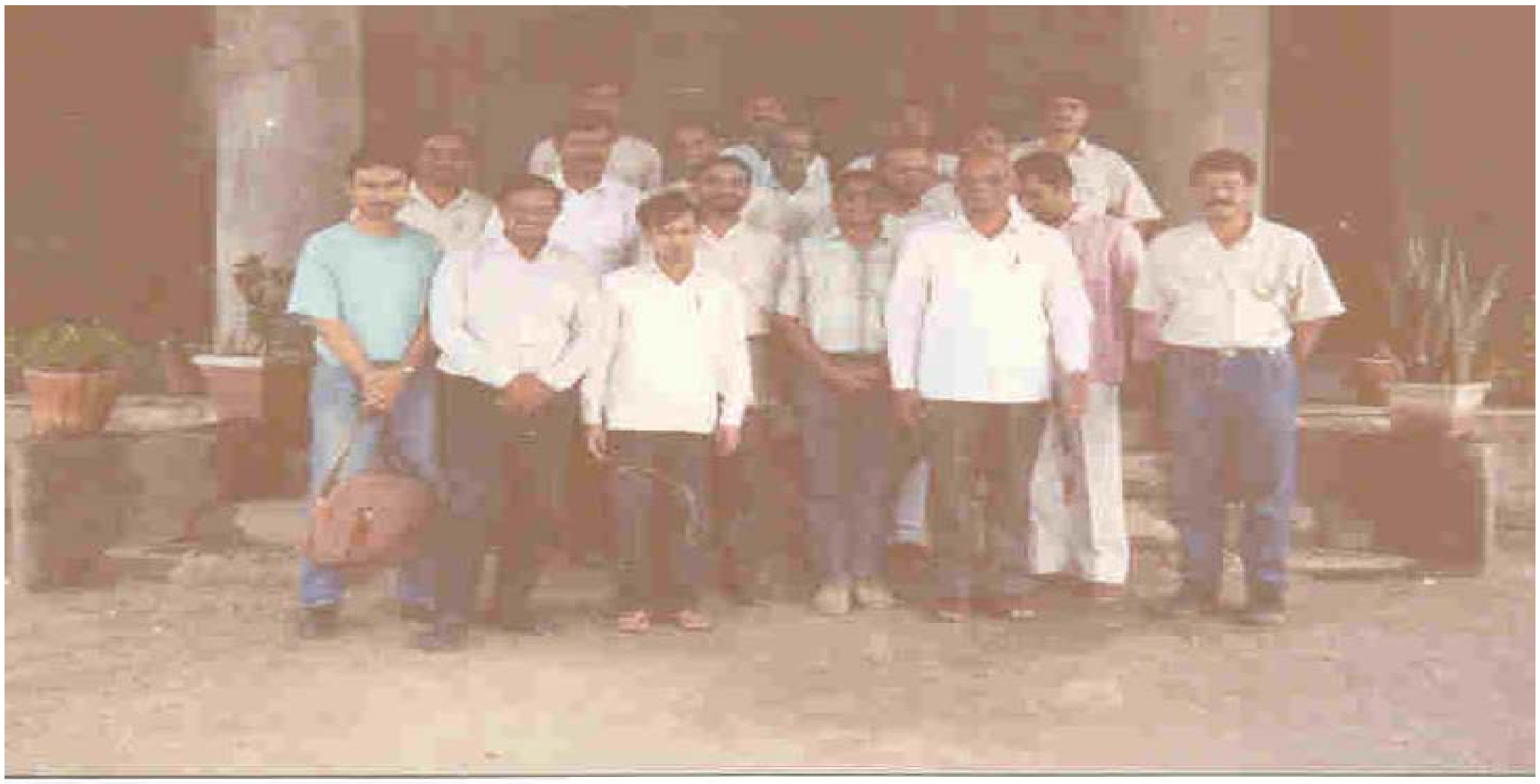

Photo 11: Mini Workshop on Photoelectric Photometry, December 2-7, 1991 where Rana is delivering his lecture

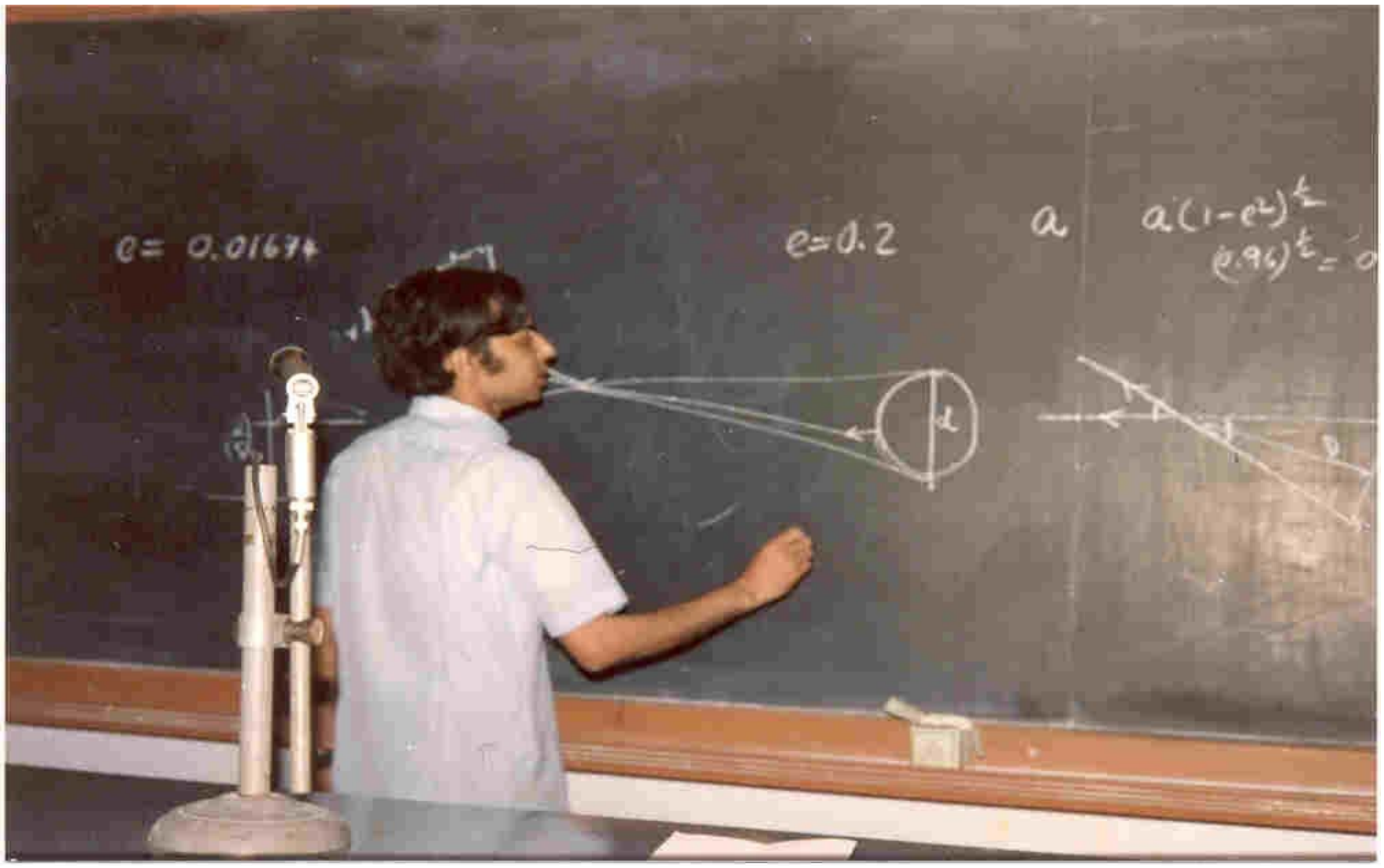

Photo 12: Mini Workshop on Photoelectric Photometry, December 2-7, 1991 where Rana is delivering his lecture

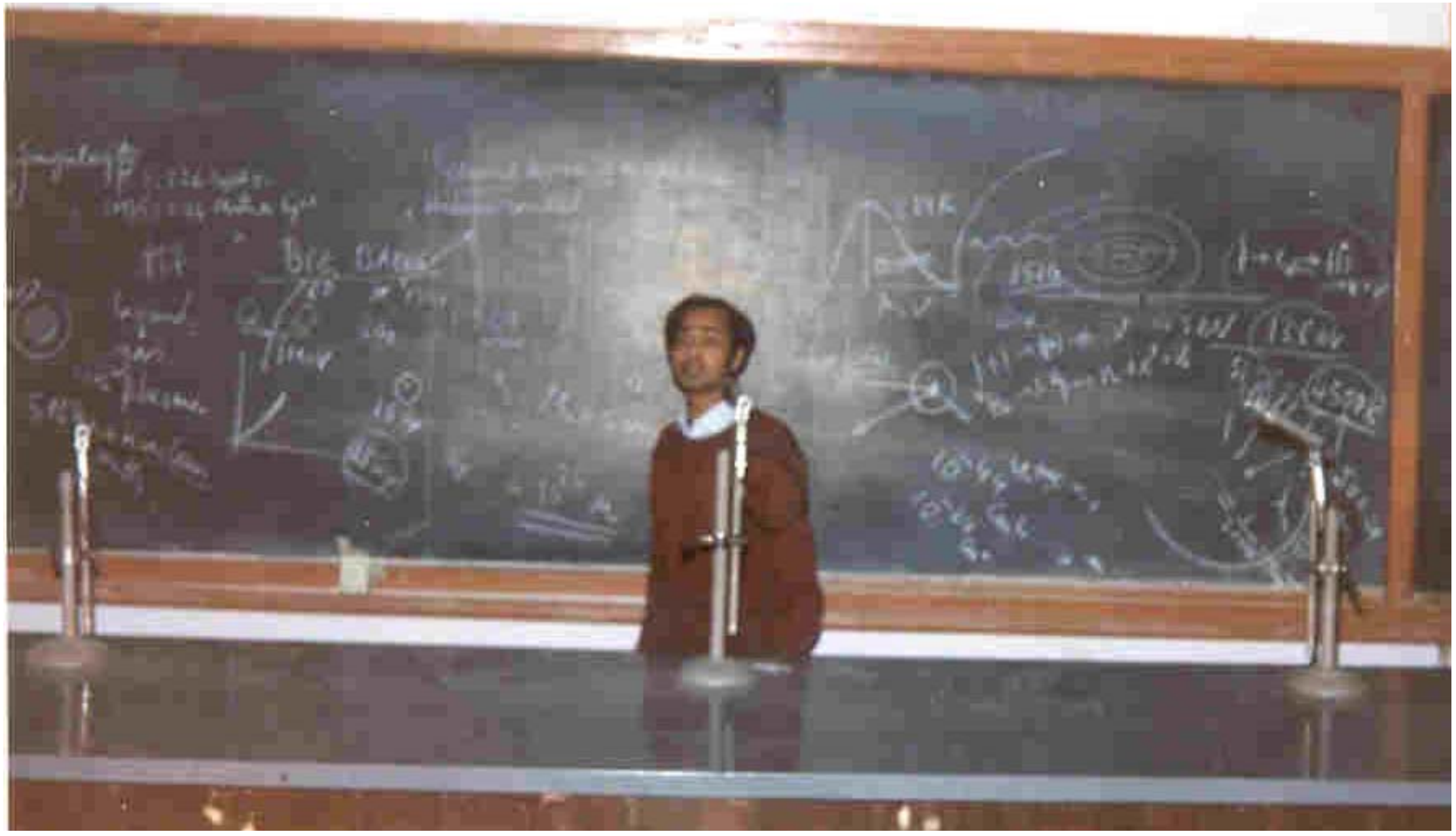

Photo 13: Refresher Course on Astronomy and Astrophysics for College and University Teachers, May 8-26, 1995 at IUCAA, Pune where Rana is seen standing beside left of the extreme right lady

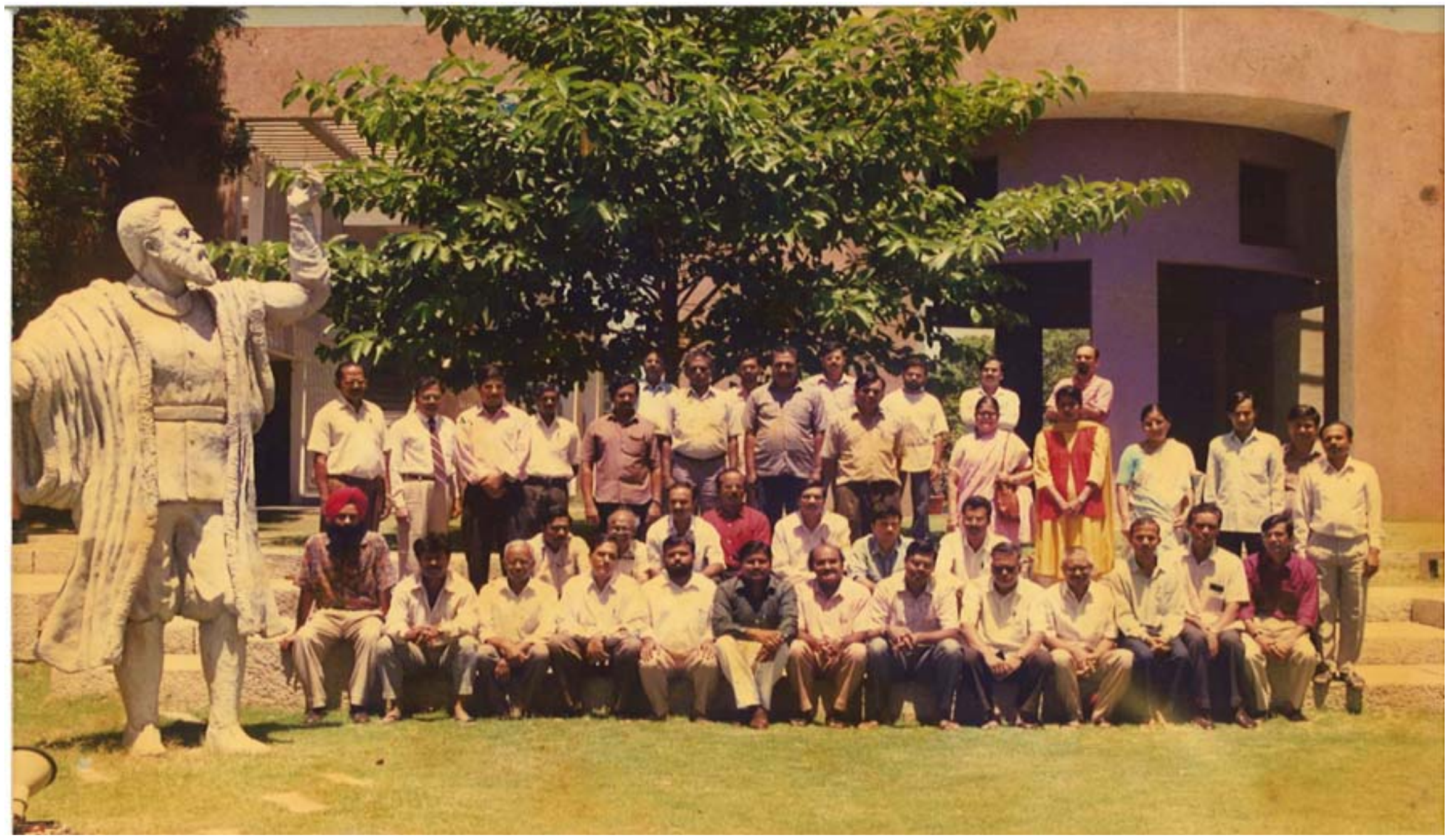

Photo 14: Workshop on Space Dynamics Celestial Mechanics at BRA Bihar University, Muzaffarpur, September 18-21, 1995 where Rana is seen seated behind the right extreme lady

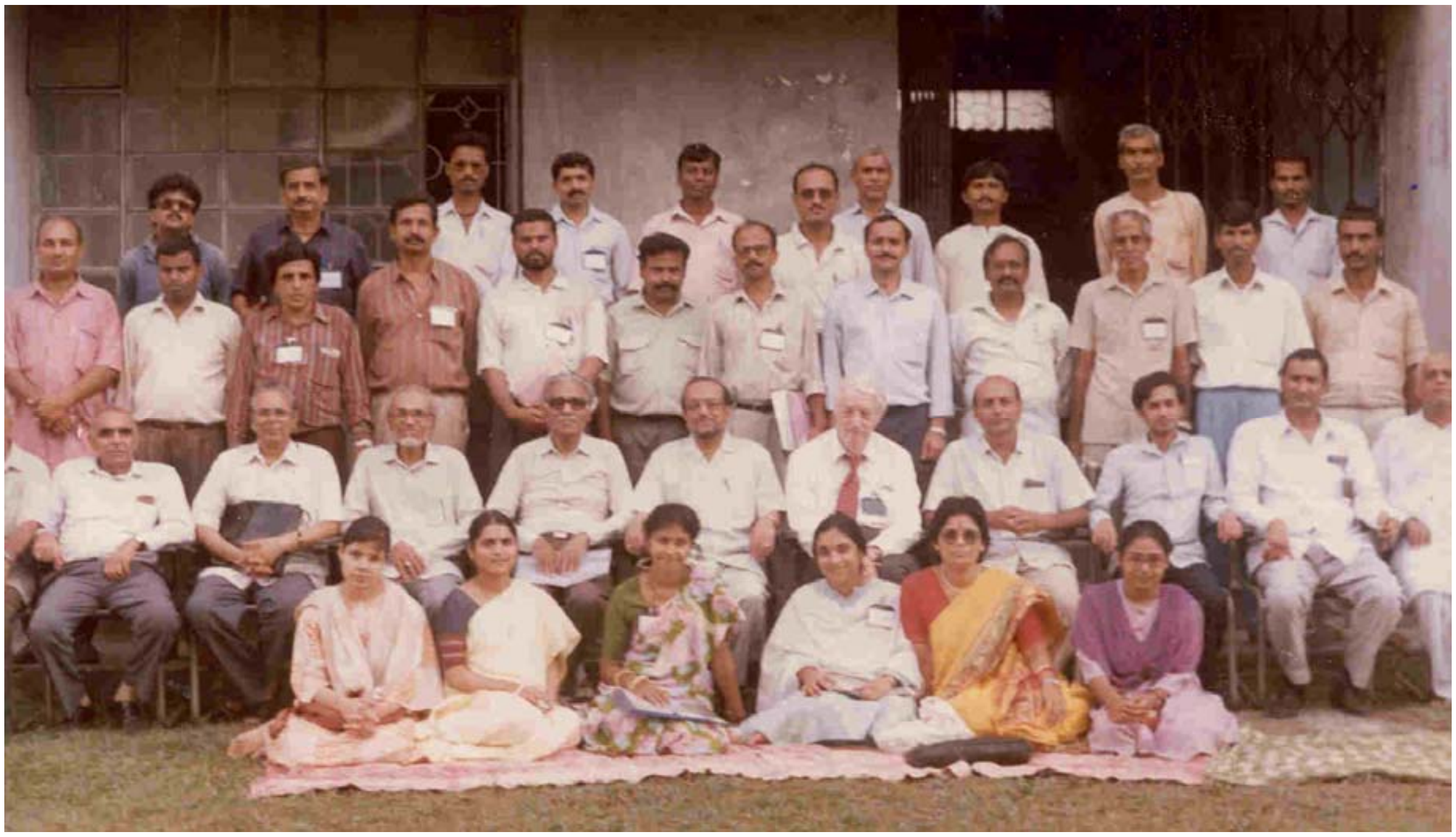

Photo 15: Emil Wolf's visit to IUCAA, Pune on 1991. From the left: Wolf, J.V. Narlikar, N. Dadhich, S.N. Tandon, N.C. Rana and G. Swaroop (the person on back is unknown).

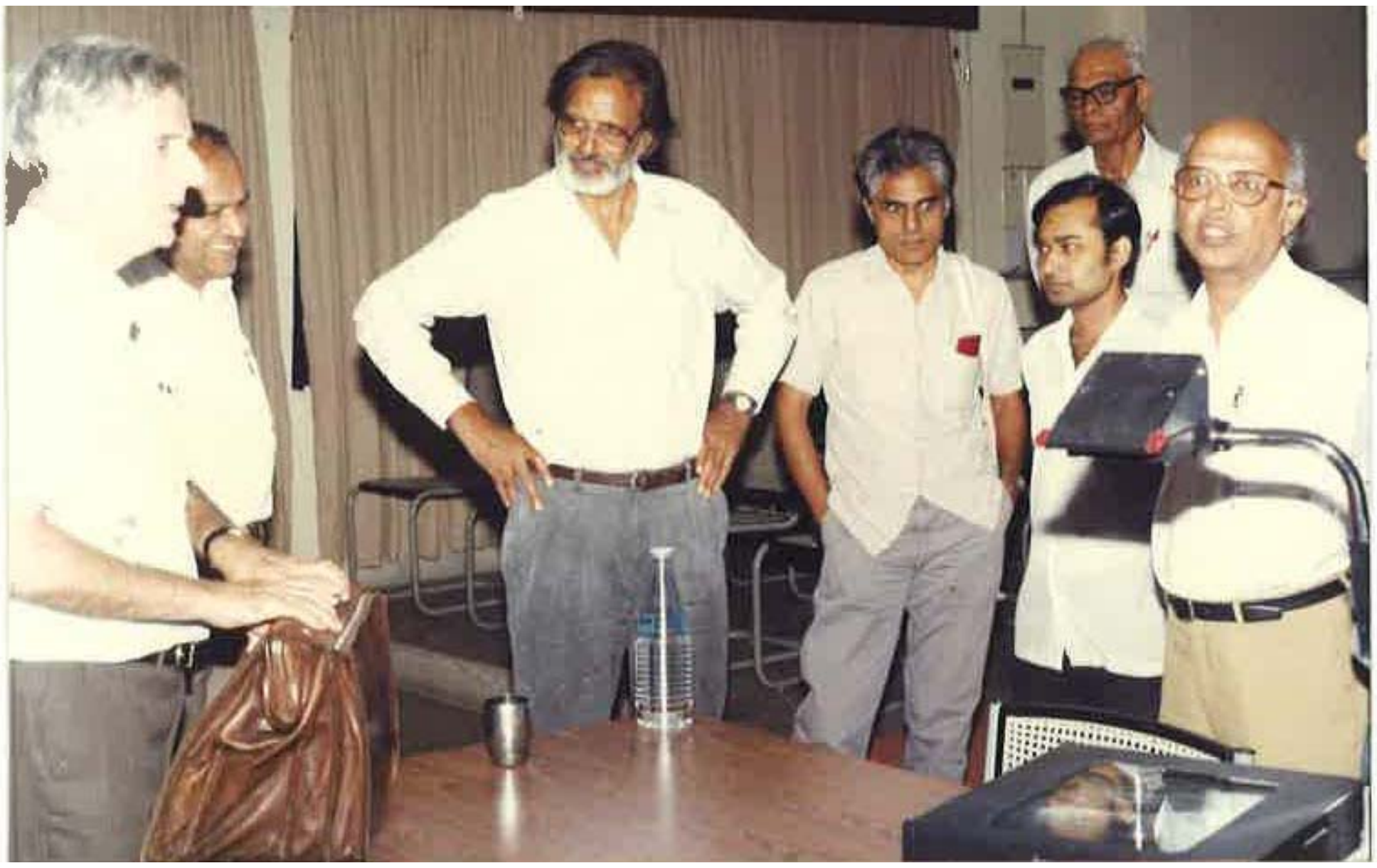

Photo 16: Fred Hoyle's visit to IUCAA, Pune on February 8, 1994. From the back row: N.C. Rana, S. Dhurandhar, A.K. Kembhavi (the present Director, IUCAA), Hoyle, J.V. Narlikar (the then Director, IUCAA), T. Padmanabhan, D.P. Duari, unknown and T.R. Shesadri.

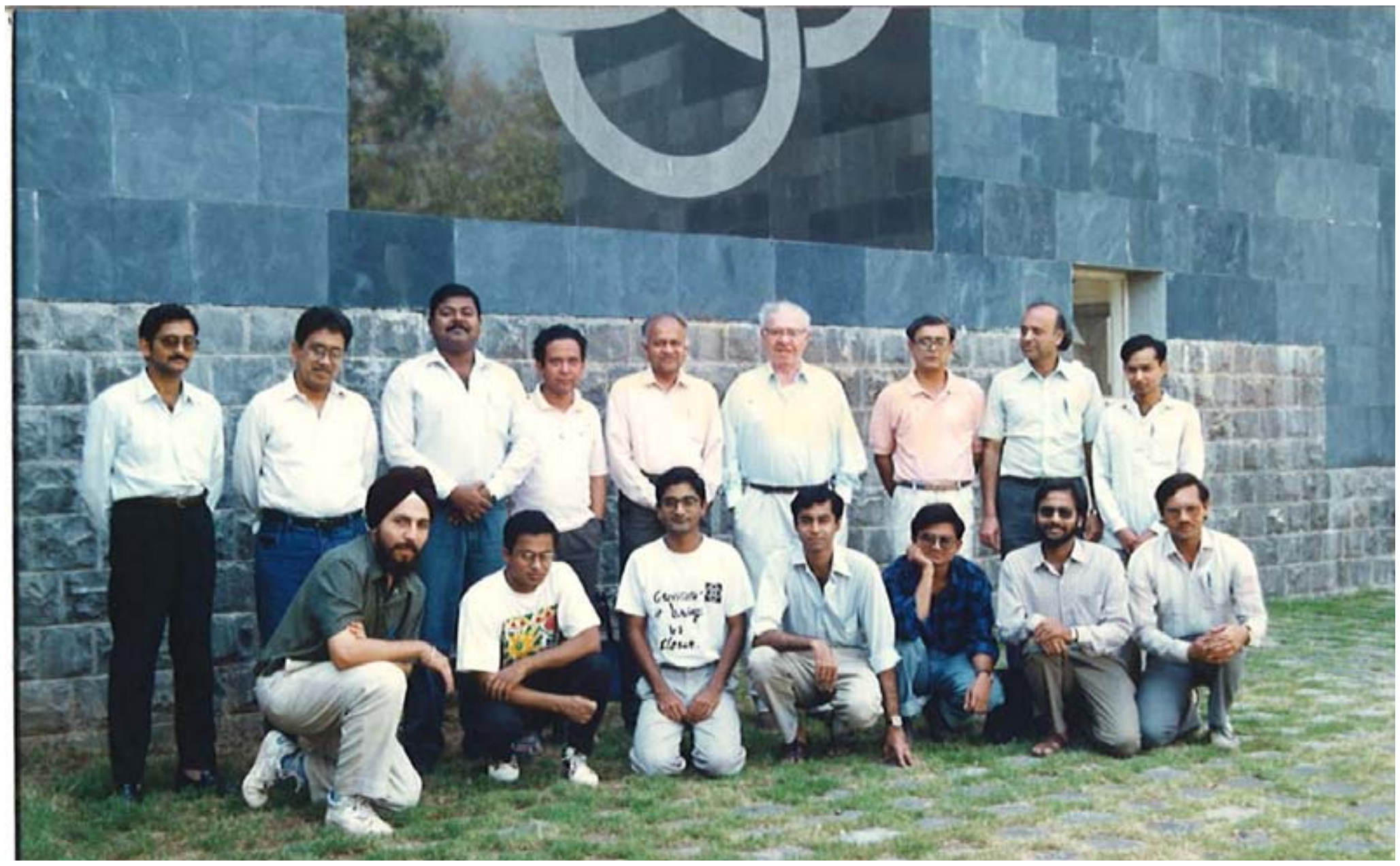

Rana also held different posts of various organizations. He was (1) Chairman, Scientific Advisory Committee, CIAA; (2) Member, Education Sub-Committee, IAU; (3) Coordinator, Popularization of Astronomical Society of India during 1993-1995; (4) Head, Science Popularization and Amateur Astronomy, IUCAA and (5) President, 'MN Lahiri Astronomical (Memorial) Trust' which he formed in loving memory of his mentor Manindra Narayan Lahiri.

One of Rana's very challenging projects of measuring the diameter of the Sun during the total solar eclipse on 24th October 1995 involved a large team of young students. It has been observed that with the knack of reaching out to the level of a layman, he was a very famous figure among amateur astronomers all over India. It was also testified that anybody who has ever had an opportunity to talk to him or hear him became infused by an almost breathless inspiration of Prof. N. C. Rana [94].

## 5. Conclusion

During 21 June to 21 July, 1996 Rana delivered lectures in Italy, Poland, England and other places of abroad. Immediately after coming back to India (perhaps on 23 July) he fell

*A very rare and informal photograph of Prof. N.C. Rana that snapped shot at his apartment in IUCAA sometime in 1994 [In courtesy of Prof. Sujan Sengupta, I.I.A., Bangalore]*

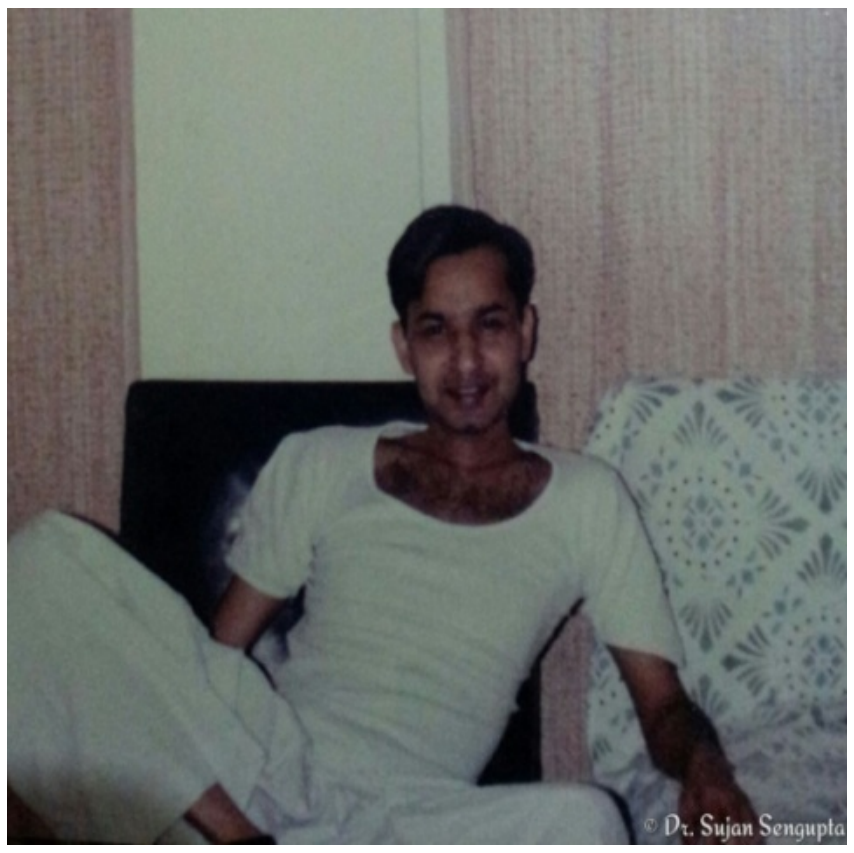

seriously ill. As his condition deteriorated rapidly, he was admitted to a nursing home in Pune. Doctors tried their best but all went in vain. At 8 am, on 22nd august, 1996 it was all over and Prof. Narayan Rana started his eternal journey. In 1997, The National Council for Science and Technology Communication (NCSTC) honored him by conferring 'National Science Popularisation Award' (posthumous) which included a citation and a sum of Rs. 1 lakh. Rana's mother Nakfuri Devi, along with a well wisher Mr. Narayan Chandra Maity, went to Delhi to receive the prize on behalf of deceased Prof. Rana.

After Rana's demise, Prof. A.K. Raychaudhuri remarked: "Narayan was the best science student that I have ever seen. I knew that he will not remain with us for a long time, but could not think that he will leave so early. If he could enjoy a longer and malady free life, then one day he would have been regarded as the greatest Bengali scientist. I can't explain my mental agony caused by his sudden demise". While narrating Rana's specialties, B.K. Pattanayak, the then Coordinator (Popularisation) of CIAA, made comment: "As a person, he was simple, friendly and humorous. As a student of science, he was competent, serious, hard working and rational. As a scientist, he was brilliant, bold and social. As a teacher, he was very affectionate and inspiring.” In the Monthly Newsletter ‘*Khagol*’ of IUCAA Rana’s Ph.D. supervisor Prof. J.V. Narlikar wrote: “Perhaps "striving" is not the right word to describe the dynamic personality of Rana. His small stature and outwardly sedate demeanour hid a highly motivated and restless human being. I discovered this, right from the times when Rana joined me as a Ph.D. student more than sixteen years ago, when we were both at the Tata Institute of Fundamental Research (TIFR), Bombay’’. According to life-long friend Raut’s version: ``Narayan was apparently calm but very sentimental. He could not but protest when anybody tried to undermine him in any way’’. This statement by Raut have later on been approved by Prof. S. Mukherjee and Prof. Mira Dey in conversation with one of the authors.

However, as a final comment we would like to put here that the life of Prof. N.C. Rana was quite dramatic having much resemblance with a comet in connection to his untimely death and the all round information about him reveals that he was absolutely a religious man – in searching of scientific truth and in his very personal pilgrimage.

## ACKNOWLEDGEMENT

The life sketch of Rana is written here by taking references from various articles of the documentary volume '*Souvenir of the First Death Anniversary Memorial Seminar on Social and Technical Aspects of Late Prof. Narayan Chandra Rana*'. Thanks are due to the authors and organizers of the said seminar. We are grateful to Prof. S. Sengupta, IIA, Bangalore, Prof. Sarbani Basu, Yale University, USA, Mr. Samir Dhurde, Incharge of Public Outreach, IUCAA, Pune and, at the last not the least, the Librarian and all the Staff of library, IUCAA, Pune for their kind assistance by providing valuable documents on Prof. N.C. Rana. Also SR is personally thankful to the authority of IUCAA, Pune for providing Visiting Associateship under which a part of this work was carried out.

## *Appendix I*

M. N. Lahiri, youngest of three brothers, was born in Rangpur of undivided Bengal (now in Bangladesh) on 7 March, 1936. His father was a *Dewan* (Finance Minister) of a local Land Lord. After partition of India, Lahiri’s family switched over to Calcutta (now Kolkata) in the year 1948 which resulted in acute financial crisis. However, Lahiri’s father was enlightened enough to provide proper education to his three sons. M.N. Lahiri was graduated from Berhampur College in 1957. After working as a school teacher for a period of two years at Chirimiri of Madhyapradesh, he finally joined Sauri Bholanath Vidyamandir in 1963 as a science teacher and spent 27 years of his life at Sauri. From his childhood, Lahiri had a knack for making small equipments and toys and made pin hole camera, telescope, star charts etc. After coming over to Sauri, he exercised this natural skill in preparing and flying kites, making small telescopes and in this way set an example before his students how one can use his (or her) leisure for nurturing creativity. He indigenously built planispheres, binoculars and a large number of telescopes of aperture ranging from 2.5 inch to 6 inch. Apart from being involved in hands on activities, Lahiri was equally efficient in writing books and articles on popular science. He wrote six books of which *‘Chander Deshe Matir Manush’ (Man in the Moon)*, *‘Ebarer Purnagras*

*Suryagrahan'* (*Total Solar Eclipse of this year* -published on the eve of total solar Eclipse 1980) and *'Dhumketu O Halleyr Dhumketu'* (*Comet and the Halley's Comet*) were published. The first one of these three books was considered for 'Rabindra Puraskar' (established after the name of Rabindranath Tagore, a Bengali Nobel Laureate in literature), but unfortunately missed final selection. One of his articles describing a novel technique he devised for the integrated finderscope and was published in *'Sky and Telescope'* magazine of USA. He used to subscribe various magazines like *'Scientific American'*, *'Sky and Telescope'* etc. Lahiri had a chronic problem in thyroid gland as a result of a neck injury which he suffered in an accident in his boyhood. While returning from England, N. C. Rana brought a 10-inch telescope for his mentor 'Manindra Sir'. In the year 1990, Lahiri fell a victim to lung cancer and spent his last days at Gurap in Hoogly district at his middle brother's residence. Finally, he succumbed to his ailment at a premature age on 5 January, 1991. N.C. Rana, the most brilliant student of M.N. Lahiri has summarized the selfless devotion of Mr. Lahiri in the following words : "Leading a passionate life for noble cause and practicing it in solitude had been his philosophy of his life; morality and service to the deserved ones had been his religion."